\documentclass[%
 aip,
 amsmath,amssymb,
 reprint,%
]{revtex4-1}

\usepackage{graphicx}
\usepackage{dcolumn}
\usepackage{bm}

\usepackage[utf8]{inputenc}
\usepackage[T1]{fontenc}
\usepackage{mathptmx}
\usepackage{etoolbox}
\usepackage{bbm}

\usepackage{amsmath,amsfonts,bm}
\usepackage{amsthm}
\usepackage{mathtools}
\usepackage{centernot}
\usepackage{algorithm}
\usepackage{algorithmic}
\usepackage{comment}
\usepackage{hyperref}

\DeclareMathOperator*{\argmax}{arg\,max}

\newcommand{\E}{\operatorname{\mathbb E}}
\newcommand{\V}{\operatorname{\mathbb V}}
\newcommand{\innermid}{\;\middle\lvert\;}

\newtheorem{remark}{Remark}
\newtheorem{corollary}{Corollary}
\newtheorem{theorem}{Theorem}

\newtheorem{definition}{Definition}
\newtheorem{assumption}{Assumption}

\allowdisplaybreaks

\makeatletter
\def\@email#1#2{%
 \endgroup
 \patchcmd{\titleblock@produce}
  {\frontmatter@RRAPformat}
  {\frontmatter@RRAPformat{\produce@RRAP{*#1\href{mailto:#2}{#2}}}\frontmatter@RRAPformat}
  {}{}
}%
\makeatother

\begin{document}

\preprint{AIP/123-QED}

\title[Hypergraphon Mean Field Games]{Hypergraphon Mean Field Games}
\author{Kai~Cui}
\affiliation{ 
Technische Universität Darmstadt, 64283 Darmstadt, Germany
}%

\author{Wasiur~R.~KhudaBukhsh}%
\affiliation{ 
University of Nottingham, Nottingham NG7 2RD, United Kingdom
}%

\author{Heinz~Koeppl}
\affiliation{%
Technische Universität Darmstadt, 64283 Darmstadt, Germany
}%
\email{heinz.koeppl@tu-darmstadt.de}

\date{\today}

\begin{abstract}
We propose an approach to modelling large-scale multi-agent dynamical systems allowing interactions among more than just pairs of agents using the theory of mean field games and the notion of hypergraphons, which are obtained as limits of large hypergraphs. To the best of our knowledge, ours is the first work on mean field games on hypergraphs. Together with an extension to a multi-layer setup, we obtain limiting descriptions for large systems of non-linear, weakly-interacting dynamical agents. On the theoretical side, we prove the well-foundedness of the resulting hypergraphon mean field game, showing both existence and approximate Nash properties. On the applied side, we extend numerical and learning algorithms to compute the hypergraphon mean field equilibria. To verify our approach empirically, we consider a social rumor spreading model, where we give agents intrinsic motivation to spread rumors to unaware agents, and an epidemics control problem.
\end{abstract}

\maketitle

\begin{quotation}
Recent developments in the field of complex systems have shown that real-world multi-agent systems are often not restricted to pairwise interactions, bringing to light the need for tractable models allowing higher-order interactions. At the same time, the complexity of analysis of  large-scale multi-agent systems on graphs remains an issue even without considering higher-order interactions. An increasingly popular and tractable approach of analysis is the theory of mean field games. We combine mean field games with higher-order structure by means of hypergraphons, a limiting description of very large hypergraphs. To motivate our model, we build a theoretical foundation for the limiting system, showing that the limiting system has a solution and that it approximates finite, sufficiently large systems well. This allows us to analyze otherwise intractable, large hypergraph games with theoretical guarantees, which we verify using two examples of rumor spreading and epidemics control.
\end{quotation}

\section{\label{sec:intro}Introduction}
In the recent years, there has been a surge of interest in large-scale multi-agent dynamical systems on higher-order networks due to their great generality and practical importance e.g. in epidemiology \cite{horstmeyer2020adaptive}, opinion dynamics \cite{iacopini2019simplicial, xu2022dynamics}, network synchronization \cite{skardal2021higher, anwar2022intralayer}, neuroscience \cite{Ziegler2022Balanced}, and more. We refer the interested readers to the excellent review articles \cite{porter2020nonlinearity, battiston2020networks, bick2021higher}. In addition to providing a more realistic description of the underlying processes, such large-scale systems with higher-order interactions pose interesting control problems for the reinforcement learning and control communities \cite{zhang2021multi,qu2019exploiting, gu2021mean}. To this end, a big challenge has been to find tractable solutions \cite{lin2021multi, qu2020scalable}. 

An increasingly popular and recent approach to the tractability issue has been to use the framework of learning in mean field games (MFGs) \cite{cardaliaguet2017learning, guo2019learning, elie2020convergence, guo2020general, cui2021approximately, chen2021maximum, bonnans2021generalized, anahtarci2021learning, guo2022entropy} and their cooperative counterpart commonly known as mean field control (MFC) \cite{arabneydi2014team, pham2018bellman, gu2020q, mondal2021approximation, cui2021discrete, mondal2022can, carmona2019model}. It is important to note that here, learning refers to the classical learning -- i.e. iterative computation -- of equilibria in game theory, as opposed to e.g. reinforcement learning, see also e.g. the discussion in \citet{lauriere2022learning}. Here are also some extensive reviews on mean field games \cite{achdou2010mean, gueant2011mean, bensoussan2013mean, carmona2018probabilistic, caines2021mean}. Popularized by \citet{huang2006large} and \citet{lasry2007mean} in the context of differential games, mean field games and related approximations have since found application in a plethora of fields such as transportation and traffic control \cite{tanaka2020linearly,  cabannes2021solving, huang2021dynamic}, large-scale batch processing and scheduling systems \cite{hanif2015mean, kar2020throughput, khudabukhsh2020generalized}, peer-to-peer streaming systems \cite{KhudaBukhsh2016P2P}, malware epidemics \cite{Eshghi2016Malware}, crowd dynamics and evacuation of buildings \cite{tcheukam2016evacuation, djehiche2017evac, aurell2018mean}, as well as many other applications in economics \cite{carmona2020applications} and engineering \cite{djehiche2017mean}. Tractably finding competitive equilibria and decentralized, cooperative optimal control solutions has been the focus of many recent works \cite{gao2017control, subramanian2019reinforcement, perrin2021generalization, perolat2021scaling, vasal2021sequential, cui2022learning, lauriere2022scalable}. Since then, mean field systems have also been extended to dynamical systems on graphs, typically using the theory of large graph limits called graphons \cite{lovasz2012large,glasscock2015graphon}. The graphon mean field systems can be considered either as the limit of systems with weakly-interacting node state processes \cite{bayraktar2020graphon, cui2022learning}, or alternatively as the result of a double limit procedure where each node constitutes a large population, or `cluster' of agents, each of which interacts with each other via inter- and intra-cluster coupling. First, infinitely many nodes are considered according to the graphon, and then infinitely many agents are considered per node, see e.g. \citet{caines2018graphon, caines2019graphon}. 

In this work, we will consider the former. The goal of our work is the synthesis of dynamical systems on hypergraphs with competitive or selfish agents. Existing analysis of hypergraph mean field systems typically remains restricted to special dynamics such as epidemiological equations \cite{bodo2016sis, landry2020effect, higham2021mean} or opinion dynamics \cite{noonan2021dynamics} on sparse graphs. In contrast, our work deals with general, agent-controlled non-linear dynamics and equilibrium solutions. We build upon prior results for discrete-time, graph-based mean field systems \cite{saldi2018markov, bayraktar2020graphon, cui2022learning} and extend them to incorporate higher-order hypergraphs as well as multiple layers. 

\begin{figure*}
    \centering
    \begin{minipage}{.2\textwidth}
    \includegraphics{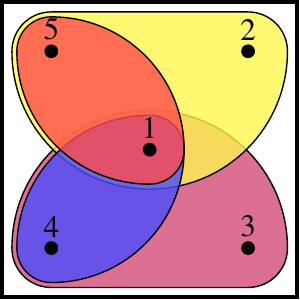}
    \end{minipage}%
    \begin{minipage}{.05\textwidth}
    \centering
    $\equiv$
    \end{minipage}%
    \begin{minipage}{.33\textwidth}
    \centering
    \includegraphics{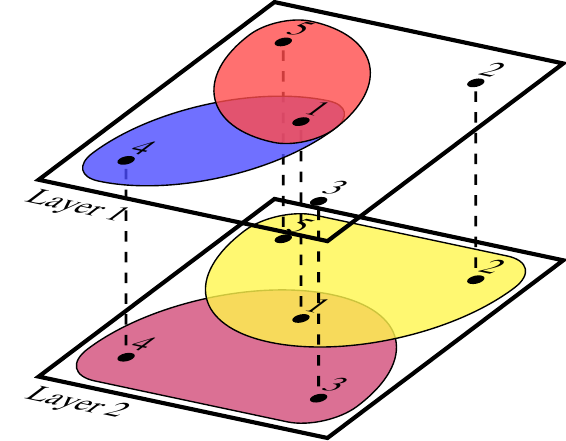}
    \end{minipage}
    \begin{minipage}{.05\textwidth}
    \centering
    $\implies$\\\vspace{2cm}
    $\implies$
    \end{minipage}%
    \begin{minipage}{.25\textwidth}
        \centering
        \includegraphics{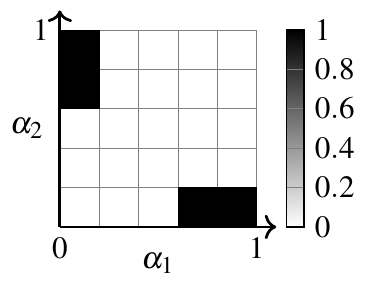}
        \includegraphics{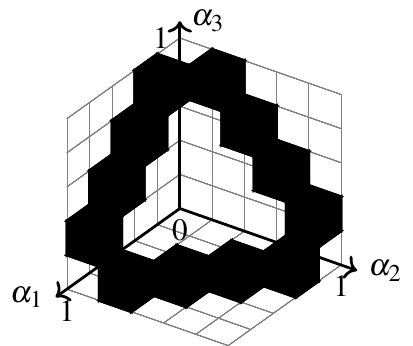}
    \end{minipage}%
    \caption{An example hypergraph $H$ is transformed into a multi-layer uniform hypergraph. On the left, a hypergraph $H$ with nodes $V[H] = \{1, \ldots, 5\}$ and hyperedges $E[H] = \{ \{1,4\}, \{1,5\}, \{1,3,4\}, \{1,2,5\} \}$ is depicted. An equivalent representation of $H$ as a $[2, 3]$-uniform hypergraph $H_\mathrm{unif}$ as well as its associated hypergraphons are given, where the first and second layers each consist of edges $\{ \{1,4\}, \{1,5\} \}$ and $3$-hyperedges $\{ \{1,3,4\}, \{1,2,5\} \}$ respectively. The associated (step-)hypergraphons $W[H^1_\mathrm{unif}]$ and $W[H^2_\mathrm{unif}]$ are given as continuous versions of the (multi-dimensional) $\{0,1\}$-valued adjacency matrices. Here, we depict only the first three coordinates for the second layer step-hypergraphon $W[H^2_\mathrm{unif}]$, given by the constant $1$ (black) or $0$ (white). Note that while each edge corresponds to two entries in the adjacency matrix of the $2$-uniform case, for the $3$-uniform case each hyperedge corresponds to six entries, resulting in the step graphon shown (bottom right).}
    \label{fig:hypergraphs}
\end{figure*} 

Our contribution can be summarized as follows: (i) To the best of our knowledge, ours is the first general mean field game-theoretical framework for non-linear dynamics on multi-layer hypergraphs. Multi-layer networks \cite{kivela2014multilayer} have proven extremely useful in many application areas including infectious disease epidemiology, where different layers could be used to describe community, household and hospital settings \cite{jacobsen2018large}. (ii) We prove the existence and the approximation properties of the proposed mean field equilibria. (iii) We propose and empirically verify algorithms for solving such hypergraphon mean field systems, and thereby obtain a tractable approach to solving and analyzing otherwise intractable Nash equilibria on multi-layer hypergraph games. The proposed framework is of great generality, extending the recently established graphon mean field games and thereby also standard mean field games (via fully-connected graphs). 

After introducing some graph-theoretical preliminaries, in Section~\ref{sec:model} we will begin by formulating the motivating mathematical dynamical model and game on hypergraphs, as well as its more tractable mean field analogue. Then, in Section~\ref{sec:theo} we will show the existence of solutions for the mean field problem as well as quantify its approximation qualities of the finite hypergraph game, building a mathematical foundation for hypergraphon mean field games. Lastly, in Section~\ref{sec:exp} we will evaluate our model numerically for an illustrative rumor spreading game, verifying our theoretical approximation results and the obtained equilibrium behavior. All of the proofs can be found in the Appendix.

\textit{Notation. \. On a discrete space $\mathbb A$, define the spaces of all (Borel) probability measures $\mathcal P(\mathbb A)$ and all sub-probability measures $\mathcal B(\mathbb A)$, equipped with the $L_1$ norm. Define the unit interval $\mathbb I \coloneqq [0, 1]$ and its $N$ equal-length subintervals $I_1^N, \ldots, I_N^N $ such that $\bigsqcup_{i=1}^N I_i^N = \mathbb I$ for any integer $N$, where $\bigsqcup$ denotes disjoint union and each $I_i^N$ includes its rightmost point $ i /N$. Denote the expectation and variance of random variables $X$ by $\E [X]$, $\V [X]$. Define the indicator function $\mathbbm 1_A(x)$ mapping to $1$ whenever $x \in A$ and $0$ otherwise. For any integer $k$, define $[k] \coloneqq \{1, \ldots, k\}$. Let $r(A,m)$ denote the set of all distinct non-empty subsets of any set $A$ with at most $m$ elements, and denote the set of all distinct non-empty, proper subsets by $r_<(A) \coloneqq r(A,|A|-1)$ as well as the set of all distinct non-empty subsets by $r(A) \coloneqq r(A,|A|)$. To keep notation simple, in the following we write $r_<[k] \coloneqq r_<([k])$, $r[k] \coloneqq r([k])$ and identify e.g. $r_<[k]$ with $[|r_<[k]|] \coloneqq \{1, \ldots, |r_<[k]|\}$ whenever helpful. Denote the set of permutations of a set $A$ as $\operatorname{Sym}(A)$. Define the space of bounded, $r_<[k]$-dimensional, symmetric functions $\operatorname{Sym}^\mathrm{ind}_<[k]$ induced by permutations of the underlying set $[k]$, i.e. any bounded function $f \colon \mathbb I^{r_<[k]} \to \mathbb R$ is in $\operatorname{Sym}^\mathrm{ind}_<[k]$ whenever $f$ is invariant to all permutations $\sigma \in \operatorname{Sym}([k])$, $f(x_1, \ldots, x_k, x_{11}, x_{12}, \ldots) = f(x_{\sigma(1)}, \ldots, x_{\sigma(k)}, x_{\sigma(1)\sigma(1)}, x_{\sigma(1)\sigma(2)}, \ldots)$. Analogously, we define spaces of such functions $\operatorname{Sym}^\mathrm{ind}_\leq[k]$ and $\operatorname{Sym}^\mathrm{ind}[k]$ over $r[k]$ and $[k]$, respectively.}

\section{\label{sec:model}Mathematical Model}

Before we formulate the stochastic dynamic hypergraph game and its limiting analogue in the following subsections, we discuss some graph-theoretical preliminaries. A (undirected) hypergraph is defined as a pair $H = (V, E)$ of a set of vertices $V$ and a set of hyperedges $E \subseteq 2^{V} \setminus \{\varnothing\}$. In contrast to edges in graphs, here hyperedges may connect an arbitrary number of vertices instead of only two. If there is no scope of confusion, we will call hyperedges  of a hypergraph just edges. Denote by $V[H]$ and $E[H]$ the vertex set and edge set of a hypergraph $H$. The maximum cardinality of all edges of a hypergraph $H$ is called its rank. A $k$-uniform hypergraph is defined as a hypergraph where all edges have cardinality $k$. A multi-layer hypergraph $H = (V, E^1, \ldots, E^D)$ with $D$ layers is obtained by allowing for multiple edge sets $E^1, \ldots, E^D \subseteq 2^{V} \setminus \{\varnothing\}$, and we analogously write $E^d[H]$ for the $d$-th set of edges of a multi-layer hypergraph $H$. We define the $d$-th sub-hypergraph $H^d$ of a multi-layer hypergraph $H$ as the hypergraph with vertex set $V[H]$ and edge set $E[H^d] = E^d[H]$.

Consider any (non-uniform) hypergraph $H$ with bounded rank $k_\mathrm{max}$. Observe the isomorphism between multi-layer uniform hypergraphs and such $H$ by splitting hyperedges of each cardinality $k \leq k_\mathrm{max}$ into their own layer. Since this procedure can be repeated for each layer of a multi-layer hypergraph, any multi-layer hypergraph is therefore equivalent to a correspondingly defined multi-layer uniform hypergraph. Hence, from here on it suffices to define and consider $[k_1, \ldots, k_D]$-uniform hypergraphs $H$ as $D$-layer hypergraphs, where each layer $d=1,\ldots,D$ is given by a $k_d$-uniform hypergraph with $k_d \leq k_\mathrm{max}$, see also Fig.~\ref{fig:hypergraphs} for a visualization. For instance, in social networks each layer could model e.g. the $k$-cliques of acquaintances formed at work, friendship at school or family relations.

To formulate the infinitely-large mean field system, we define the limiting description of sufficiently dense multilayer hypergraphs as the graphs intuitively become infinite in size, called hypergraphons \cite{elek2012measure}. Here, dense means a number of edges on the order of $O(N^2)$, where $N$ is the number of vertices, to which existing hypergraphon theory remains limited to. However, we note that an extension to more sparse models by fusing the theory of hypergraphons with $L^p$ graphons \cite{borgs2018p, borgs2019p, fabian2022learning} could be part of future work. The space of $k$-uniform hypergraphons $\mathbb W_k$ is now defined as the space of all bounded and symmetric functions $W \in \operatorname{Sym}^\mathrm{ind}_<[k], W \colon \mathbb I^{r_<[k]} \to \mathbb I$ that are measurable. We equip $\operatorname{Sym}^\mathrm{ind}_<[k]$ with the cut (semi-)norm $\lVert \cdot \rVert_{\square^{k-1}}$ proposed by \citet{zhao2015hypergraph}, defined by
\begin{align} \label{eq:cutconv}
    &\lVert W \rVert_{\square^{k-1}} \coloneqq \sup_{\substack{u_i \colon \mathbb I^{r[k-1]} \to \mathbb I, \\u_i \in \operatorname{Sym}^\mathrm{ind}_\leq[k-1]}} \left| \int_{\mathbb I^{r_<[k]}} W(\alpha) \prod_{i=1}^k u_i(\alpha_{r([k] \setminus \{i\}})) \, \mathrm d\alpha \right|, 
\end{align}
which (see e.g. \citet[Lemma 8.10]{lovasz2012large}) coincides with the standard graphon case for $k=2$,
\begin{align}
    \lVert W \rVert_\square = \sup_{f,g \colon \mathbb I \to \mathbb I} \left| \int_{\mathbb I^2} W(\alpha, \beta) f(\alpha) g(\beta) \, \mathrm d(\alpha, \beta) \right|.
\end{align}

\begin{figure}
    \centering
    \begin{minipage}{.4\linewidth}
        \centering
        \includegraphics{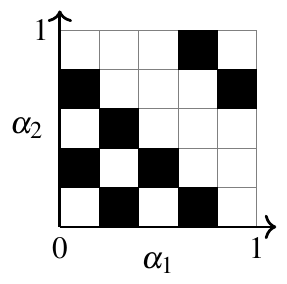}
    \end{minipage}
    \begin{minipage}{.02\linewidth}
    \centering
    $\to$
    \end{minipage}%
    \begin{minipage}{.52\linewidth}
        \centering
        \includegraphics{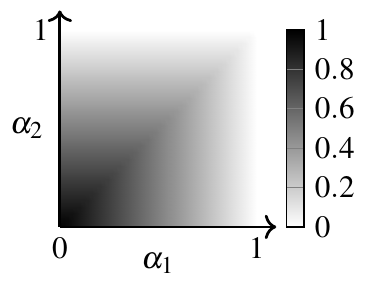}
    \end{minipage}
    \caption{Visualization of the convergence of $2$-dimensional step-graphons to the uniform attachment graphon $W_\mathrm{unif}(\alpha_1, \alpha_2) = 1 - \max(\alpha_1, \alpha_2)$.}
    \label{fig:hypergraphonconv}
\end{figure}

To analytically connect $k$-uniform hypergraphs to hypergraphons, we define the step-hypergraphons of any $k$-uniform hypergraph $H$ as
\begin{align}
    W[H](\boldsymbol \alpha) = \sum_{\mathbf m \in [N]^{k}} \mathbbm 1_{E[H]}(\mathbf m) \cdot \prod_{i \in [k]} \mathbbm 1_{I_{m_i}^N}(\alpha_{i}).
\end{align}

For motivation, note that for any sequence of graphs with converging homomorphism densities, equivalently the step graphons converge in the cut norm to the limiting graphon, and their limiting homomorphism densities can be described by the limiting graphon \cite{lovasz2012large}. Similarly, cut-norm convergence for the more general uniform hypergraphs at least implies the convergence of hypergraph homomorphism densities \cite{zhao2015hypergraph}. Accordingly, we assume hypergraph convergence in each layer of a given sequence of $[k_1, \ldots, k_D]$-uniform hypergraphs $(H_N)_{N \in \mathbb N}$ via convergence of their step-hypergraphons $W^N_d \coloneqq W[H^d_N]$ to a limiting hypergraphon $W_d \in \mathbb W_{k_d}$ in the cut norm as visualized in Fig.~\ref{fig:hypergraphonconv}, similar as in standard graphon mean field systems \cite{bayraktar2020graphon, cui2022learning}.

\begin{assumption} \label{ass:W}
The sequence of step-hypergraphons $W^N \coloneqq (W^N_d)_{d \in [D]}$ converges on each layer in cut norm $\left\Vert \cdot \right\Vert_{\square}$ to some hypergraphons $W \coloneqq (W_d)_{d \in [D]} \in \bigtimes_{d \in [D]} \mathbb W_{k_d}$, i.e.
\begin{align} \label{eq:Wconv}
    \left\Vert W^N_d - W_d \right\Vert_{\square} \to 0, \quad \forall d \in [D].
\end{align}
\end{assumption}

\subsection{Finite Hypergraph Game} 
In this subsection, we will formulate a dynamical model on hypergraphs where each node is understood as an agent that is influenced by the state distribution of all of its neighbors, according to some time-varying dynamics. Furthermore, each agent is expected to selfishly optimize its own objective, which gives rise to Nash equilibria as the solution of interest. 

Consider a $[k_1, \ldots, k_D]$-uniform hypergraph and let $\mathbb T$ be the time index set, either $\mathbb T = \{0, 1, \ldots, T-1\}$ or $\mathbb T = \mathbb N_0 \coloneqq \{0, 1, 2, \ldots \}$. We define $N$ agents $i \in [N]$ each endowed with local states $X^i_t$ and actions $U^i_t$ from a finite state space $\mathbb X$ and finite action space $\mathbb U$, respectively. Here, $\mathbb X$ and $\mathbb U$ are assumed finite for technical reasons, though we believe that results could be extended to more general spaces in the future. States have an initial distribution $X^i_0 \sim \mu_0 \in \mathcal P(\mathbb X)$. For all times $t \in \mathbb T$ and agents $i \in [N]$, their actions are random variables following the law
\begin{align}
    U^i_t \sim \pi^i_t(\cdot \mid X^i_t),
\end{align}
with policy (i.e. probability distribution over actions) $\pi^i \in \Pi \coloneqq \mathcal P(\mathbb U)^{\mathbb T \times \mathbb X}$, that, for each node $i$, depends on the $i$-th state at time $t$. Then, the states are random variables following the law
\begin{align}
    X^i_{t+1} \sim P_t(\cdot \mid X^i_t, U^i_t, \nu^{N,i}_t), 
\end{align}
with transition kernels $P_t \colon \mathbb X \times \mathbb U \times \mathcal B(\mathbb X) \to \mathcal P(\mathbb X)$ that, for each node $i$, depends on the $i$-th state and action at time $t$, and $\nu^{N,i}_t$. Here, the $\bigtimes_{d=1}^{D} \mathcal P(\mathbb X^{k_d-1})$-valued multi-layer empirical neighborhood mean field $\nu^{N,i}_t$ is defined as
\begin{align} \label{eq:empneighbormf}
    \nu^{N,i}_{t,d} \coloneqq \frac 1 {N^{k_d-1}} \sum_{\mathbf m \in [N]^{k_d-1}} \mathbbm 1_{E^d[H_N]}(\mathbf m \cup i) \delta_{\bigtimes_{j\neq i} X^{m_j}_t}, 
\end{align}
in its $d$-th layer, consisting of the unnormalized state distributions of an agent $i$'s neighbors on each layer. In other words, the state dynamics of an agent depend only on the states of nodes in their immediate neighborhood and can be influenced by the agent via its actions $U^i_t$. 

For example, in an epidemics spread scenario, the states of each agent could model their infection status, while the actions of an agent could be to take protective measures. As a result, each agent will randomly become infected with probability depending on how many neighboring agents are infected and whether the agent is taking protective measures.

The cost functions $R_t \colon \mathbb X \times \mathbb U \times \mathcal B(\mathbb X) \to \mathbb R$ with discount factor $\gamma \in (0,1)$ or in the finite horizon case $\gamma \in (0, 1]$ define the objective function for the $i$-th agent
\begin{align}
    J_i^N(\pi^1, \ldots, \pi^N) \coloneqq \E \left[ \sum_{t \in \mathbb T} \gamma^t R_t(X_{t}^i, U_{t}^i, \nu^{N,i}_t) \right],
\end{align} 
which can describe also e.g. random rewards $R^i_t$ that are conditionally independent given $X_{t}^i, U_{t}^i, \nu^{N,i}_t$ by the law of total expectation and taking the conditional expectation, $R_t(X_{t}^i, U_{t}^i, \nu^{N,i}_t) \equiv \mathbb E \left[ R^i_t \innermid X_{t}^i, U_{t}^i, \nu^{N,i}_t \right]$.

Our goal is now to find Nash equilibria, i.e. stable policies where no agent can singlehandedly deviate and improve their own objective. Note that finding Nash equilibria in games such as the above is difficult, since a) even existence of Nash equilibria under the above, decentralized information structure of policies is hard to show, and b) computation of the Nash equilibria fails due to both curse of dimensionality under full observability and general complexity of computing Nash equilibria \cite{daskalakis2009complexity}, see also \citet{saldi2018markov} and the discussion therein. 

Thus, in the finite game, we are interested in finding the following weaker notion of approximate equilibria \cite{elie2020convergence, cui2022learning}, where a negligible fraction of agents that remains insignificant to all other agents may remain suboptimal.

\begin{definition}
An $(\varepsilon, p)$-Nash equilibrium for $\varepsilon, p > 0$ is defined as a tuple of policies $(\pi^1, \ldots, \pi^N) \in \Pi^N$ such that for any $i \in \mathbb J^N$, we have
\begin{multline}
    J_i^N({\pi^1}, \ldots, {\pi^N}) \\ 
    \geq \sup_{\pi \in \Pi} J_i^N({\pi^1}, \ldots, {\pi^{i-1}}, \pi, {\pi^{i+1}}, \ldots, {\pi^N}) - \varepsilon
\end{multline}
for some set $\mathbb J^N \subseteq [N]$ of at least $\left\lfloor (1-p)N \right\rfloor$ agents.
\end{definition}

While it may seem excessive to reduce to approximate optimality limited to a fraction of the agents, it is always possible under Assumption~\ref{ass:W} for a finite number of agents to deviate arbitrarily from the limiting system description. Therefore, under our assumptions it is only possible to obtain an approximate equilibrium solution for almost all agents via the mean field formulation. Although we could make stronger assumptions on the mode of convergence for hypergraphons, such a concept of convergence would be difficult to motivate from a graph theoretical perspective. Therefore, we restrict ourselves to the cut-norm convergence \cite{zhao2015hypergraph} and the above solution concept.

\subsection{Hypergraphon Mean Field Game} 
Next, we will formally let $N \to \infty$ and obtain a more tractable, reduced model consisting of any single representative agent and the distribution of agent states, the so-called mean field. 

To analyze the case $N \to \infty$ however, we first introduce some preliminary definitions. We define the space of mean fields $\mathbb M \subseteq \mathcal P(\mathbb X)^{\mathbb T \times \mathbb I}$ such that $\boldsymbol \mu \in \mathbb M$ whenever $\alpha \mapsto \mu^\alpha_t(x)$ is measurable for all $t \in \mathbb T$, $x \in \mathbb X$. Intuitively, a mean field is the distribution of states each of the infinitely many agents in $\mathbb I$ is in. Analogously, the space of policies $\boldsymbol \Pi \subseteq \Pi^{\mathbb I}$ is given by policies $\pi \in \Pi^{\mathbb I}$ where $\alpha \mapsto \pi^\alpha_t(u \mid x)$ is measurable for any $t \in \mathbb T, x \in \mathbb X, u \in \mathbb U$. Intuitively, $\pi \in \Pi^{\mathbb I}$ defines the behavior for each agent $\alpha \in \mathbb I$. For any $f \colon \mathbb X \times \mathbb I \to \mathbb R$ and state marginal ensemble $\boldsymbol \mu \in \mathcal P(\mathbb X)^{\mathbb I}$, define $$\boldsymbol \mu(f) \coloneqq \int_{\mathbb I} \sum_{x \in \mathbb X} f(x, \alpha) \mu^\alpha(x) \, \mathrm d\alpha .$$

In the limit of $N \to \infty$, assuming that all agents follow a policy $\boldsymbol \Pi \subseteq \Pi^{\mathbb I}$, we obtain infinitely many agents $\alpha \in \mathbb I$, for each of whom we define the limiting hypergraphon mean field dynamics analogously to the finite hypergraph game. 

The agent states have the initial distribution $X^\alpha_0 \sim \mu_0 \in \mathcal P(\mathbb X)$. For all times $t \in \mathbb T$ and agents $\alpha \in \mathbb I$, their actions will be random variables following the law
\begin{align}
    U^\alpha_t \sim \pi^\alpha_t(\cdot \mid X^\alpha_t),
\end{align}
under the policy $\pi^\alpha \in \Pi$, while their states follow the law 
\begin{align}
    X^\alpha_{t+1} \sim P_t(\cdot \mid X^\alpha_t, U^\alpha_t, \nu^\alpha_t), 
\end{align}
with the limiting, now deterministic neighborhood mean field $\nu^\alpha_t \in \bigtimes_{d=1}^{D} \mathcal P(\mathbb X^{k_d-1})$. Informally, by a law of large numbers, we have replaced the distribution of finitely many neighbor states by the limiting mean field distribution $\nu^\alpha_t$. The $d$-th component of this mean field is given by
\begin{align} \label{eq:neighbormf}
    \nu^\alpha_{t,d} \coloneqq \int_{\mathbb I^{r_<[k_d] \setminus \{1\}}} W_d(\alpha, \boldsymbol \beta) \prod_{j=1}^{k_d-1} \mu^{\beta_j}_t(\cdot) \, \mathrm d\boldsymbol \beta,
\end{align}
where for readability, $(\cdot)$ denotes separate coordinates of the input (the order does not matter due to symmetry). In other words, the $d$-layer neighborhood mean field distributions are functions 
\begin{align}
    (x_1, \ldots, x_{k_d-1}) \mapsto \int_{\mathbb I^{r_<[k_d] \setminus \{1\}}} W_d(\alpha, \boldsymbol \beta) \prod_{j=1}^{k_d-1} \mu^{\beta_j}_t(x_j) \, \mathrm d\boldsymbol \beta
\end{align}
that give the probability of random neighbors of a shared hyperedge on layer $d$ to be in states $(x_1, \ldots, x_{k_d-1}) \in \mathbb X^{k_d-1}$

Note that the same, shared $\alpha \in \mathbb I$ is used for all $D$ layers, i.e. all layer neighborhood distributions of agents jointly converge to the limiting descriptions $\nu^\alpha_t$. This makes sense, since by Assumption~\ref{ass:W}, we assume that the agents are already ordered such that the corresponding step-hypergraphons converge to the limiting hypergraphon in cut norm on all layers jointly.

Finally, the objective will be given by
\begin{align}
    J^{\boldsymbol \mu}_{\alpha}(\pi^\alpha) \coloneqq \E \left[ \sum_{t \in \mathbb T} \gamma^t R_t(X_{t}^\alpha, U_{t}^\alpha, \nu^\alpha_t) \right],
\end{align}
which leads to the mean field counterpart of Nash equilibria. Informally, a mean field (Nash) equilibrium is given by a `consistent' tuple of policy and mean field, such that the policy is optimal under the mean field and the mean field is generated by the policy. As a result, if all agents follow the policy, they will be optimal under the generated mean field, leading to a Nash equilibrium.

More formally, we define the maps $\Phi \colon \mathbb M \to 2^{\boldsymbol \Pi}$ mapping from fixed mean field $\boldsymbol \mu \in \mathbb M$ to all optimal policies $\boldsymbol \pi \in \boldsymbol \Pi \colon \forall \alpha \in \mathbb I \colon \pi^\alpha \in \argmax_{\tilde \pi} J^{\boldsymbol \mu}_{\alpha}(\tilde \pi)$ and similarly $\Psi \colon \boldsymbol \Pi \to \mathbb M$ mapping from policy $\boldsymbol \pi \in \boldsymbol \Pi$ to its induced mean field $\boldsymbol \mu \in \mathbb M$ such that for all $\alpha \in \mathbb I$, $t \in \mathbb T$ we have the initial distribution $\mu^\alpha_0 = \mu_0$ and mean field evolution
\begin{align} \label{eq:mfevol}
    \mu^\alpha_{t+1} = \int_{\mathbb X} \int_{\mathbb U} P(x, u, \nu^\alpha_t) \pi^\alpha_t(\mathrm du \mid x) \mu^\alpha_t(\mathrm dx).
\end{align}

\begin{definition}
A Hypergraphon Mean Field Equilibrium (HMFE) is a pair $(\boldsymbol \pi, \boldsymbol \mu) \in \boldsymbol \Pi \times \mathbb M$ such that $\boldsymbol \pi \in \Phi(\boldsymbol \mu)$ and $\boldsymbol \mu = \Psi(\boldsymbol \pi)$.
\end{definition}

Importantly, the mean field game will be motivated rigorously in the following, and its computational complexity is independent of the number of agents. Instead, the complexity of the problem will scale with the size of agent state and action spaces $\mathbb X$, $\mathbb U$ and the considered time horizon in case of a finite horizon cost function, since we will solve for equilibria by repeatedly (i) computing optimal policies for discrete Markov decision processes \cite{puterman2014markov} $\pi^\alpha \in \argmax_{\tilde \pi} J^{\boldsymbol \mu}_{\alpha}(\tilde \pi)$, and (ii) solving the mean field evolution equations \eqref{eq:mfevol}. In particular, mean field equilibria are guaranteed to exist, and the corresponding equilibrium policy will provide an equilibrium for large finite systems. 

To obtain meaningful results, we need a standard continuity assumption (e.g. \citet{bayraktar2020graphon}), since otherwise weak interaction is not guaranteed: Without continuity, a change of behavior in only one of many agents could otherwise cause arbitrarily large changes in the dynamics or rewards.

\begin{assumption} \label{ass:Lip}
Let $R_t$, $P_t$, $W$ each be Lipschitz continuous with Lipschitz constants $L_R, L_P, L_W > 0$.
\end{assumption}

\begin{remark}
For all but Theorem~\ref{thm:existence}, we may alternatively let $W$ be Lipschitz on finitely many disjoint hyperrectangles, i.e. let there be disjoint intervals $\{I_1, \ldots,I_Q\}$, $\cup_{i} I_i = \mathbb I$ such that $\forall i \in \{1, \ldots, Q\}$, $\forall \alpha, \tilde \alpha \in I_i$, $\forall d \in [D]$, $\forall \beta \in \mathbb I^{r_<[k_d] \setminus \{1\}}$ we have
\begin{align} \label{eq:blockwiseLip}
    |W_d(\alpha, \beta) - W_d(\tilde \alpha, \beta)| \leq L_W \left| \alpha - \tilde \alpha \right|.
\end{align}
\end{remark}

\begin{remark} \label{remark:actiondep}
Note that our model is quite general: In particular, it is also possible to model dynamics and rewards dependent on the state-action distributions instead of only state distributions, replacing $\delta_{\bigtimes_{j\neq i} X^{m_j}_t}$ by $\delta_{\bigtimes_{j\neq i} (X^{m_j}_t, U^{m_j}_t)}$ in \eqref{eq:empneighbormf}. This can be done by reformulating any problem as follows. Assume a problem with state and action spaces ${\mathbb X}$, ${\mathbb U}$ and dependence of rewards and transitions on joint state-action distributions. We can rewrite the problem as a new problem with new state space ${\mathbb X} \cup \left( {\mathbb X} \times {\mathbb U} \right)$, where in the new problem, each two decision epochs $t$, $t+1$ correspond to a single original decision epoch, where in the first step $t$ we transition deterministically from $X^{m_j}_t$ to $(X^{m_j}_t, U^{m_j}_t)$ for the taken action $U^{m_j}_t$, while in the second step $t+1$ we transition and compute rewards according to the original system, ignoring any second actions taken. Choosing the square root of the discount factor and normalizing rewards will give a problem in our form that is equivalent to the original problem.
\end{remark}

\section{\label{sec:theo}Theoretical Results}
In this section, we rigorously motivate the mean field formulation by providing existence and approximation results of an HMFE. Essentially, HMFEs are guaranteed to exist and will give approximate Nash equilibria in finite hypergraph games with many agents. The reader interested primarily in applications may skip this section. 

We lift the empirical distributions and policies to the continuous domain $\mathbb I$, i.e. for any $(\pi^1, \ldots, \pi^N) \in \Pi^N$ we define the step policy $\boldsymbol \pi^N \in \boldsymbol \Pi$ and the step empirical measures $\boldsymbol \mu^N \in \mathbb M$ by
\begin{align}
    \pi^{N,\alpha}_t &\coloneqq \sum_{i \in [N]} \mathbbm 1_{I_i^N}(\alpha) \cdot \pi^i_t, \quad \forall (\alpha, t) \in \mathbb I \times \mathbb T, \\ 
    \mu^{N,\alpha}_t &\coloneqq \sum_{i \in [N]} \mathbbm 1_{I_i^N}(\alpha) \cdot \delta_{X^j_t}, \quad \forall (\alpha, t) \in \mathbb I \times \mathbb T .
\end{align} 
Proofs for the results to follow can be found in the Appendix and are at least structurally similar to proofs in \citet{cui2022learning}, though they contain a number of additional considerations that we highlight in the appendix.

\subsection{Existence of equilibria}
First, we show that there exists an HMFE. We do this by rewriting the problem in a more convenient form as done in \citet{cui2022learning}. Consider an equivalent, more standard mean field game with states $(\alpha_t, \tilde X_t)$, i.e. we integrate the graphon indices $\alpha$ into the state. The newfound states follow the initial distribution $\tilde X_0 \sim \mu_0$, $\alpha_0 \sim \mathrm{Unif}(\mathbb I)$. Then, the actions and original state transitions follow as before, while the $\alpha_t$ part of the state remains fixed at all times, i.e.
\begin{align} \label{eq:inducMDP}
    \tilde U_t &\sim \tilde \pi_t(\cdot \mid \tilde X_t, \alpha_t), \nonumber\\ \tilde X_{t+1} &\sim P_t(\cdot \mid \tilde X_t, \tilde U_t, \tilde \nu_t ), \quad \alpha_{t+1} = \alpha_t
\end{align}
where we used the standard (non-graphical) mean field $\tilde \mu_t \in \mathcal P(\mathbb X \times \mathbb I)$ (cf. \citet{saldi2018markov}) and let
\begin{align}
    \tilde \nu_{t,d} &= \int_{\mathbb I^{r_<[k_d] \setminus \{1\}}} W_d(\alpha_t, \boldsymbol \beta) \prod_{j=1}^{k_d-1} \tilde \mu_t(\cdot, \beta_j) \, \mathrm d\boldsymbol \beta,
\end{align}
Using existing results for mean field games\cite{saldi2018markov}, we obtain existence of a potentially non-unique HMFE. 

\begin{theorem} \label{thm:existence}
Under Assumption~\ref{ass:Lip}, there exists a HMFE $(\boldsymbol \pi, \boldsymbol \mu) \in \boldsymbol \Pi \times \mathbb M$.
\end{theorem}

For uniqueness results, we refer to existing results such as the classical monotonicity condition \cite{lasry2007mean, perolat2021scaling}. However, using existing theory will not analyze the finite hypergraph structure and instead directly uses the limiting hypergraphons. In the following, we thus show also that the finite hypergraph games are indeed approximated well. 

\subsection{Approximation properties}
Next, we will show that the finite hypergraph game and its dynamics are well-approximated by the hypergraphon mean field game, which implies that the HMFE solution of the hypergraphon mean field game will give us the desired $(\varepsilon, p)$-Nash equilibrium in large finite hypergraph games. 

To begin, we define and obtain finite $N$-agent system equilibria from an HMFE via the policy sharing map $\mathrm{Id}_N(\boldsymbol \pi) \coloneqq (\pi^1, \ldots, \pi^N) \in \Pi^N$, i.e. $\mathrm{Id}_N$ is defined such that each agent will act according to its position $\alpha$ on the hypergraphon,
\begin{align}
    \pi^i_t(u \mid x) \coloneqq \pi^{\frac i N}_t(u \mid x), \quad \forall (i, t, x, u) \in [N] \times \mathbb T \times \mathbb X \times \mathbb U.
\end{align} 

Now consider $(i,\hat \pi)$-deviated policy tuples where the $i$-th agent deviates from an equilibrium policy tuple to its own policy $\hat \pi$, i.e. policy tuples $(\pi^1, \ldots, \pi^{i-1}, \hat \pi, \pi^{i+1}, \ldots, \pi^N)$. Note that this includes the deviation-free case as a special case. In order to obtain a $(\varepsilon, p)$-Nash equilibrium, we must show that for almost all $i$ and policies $\hat \pi$, the $(i,\hat \pi)$-deviated policy tuple will be approximately described by the interaction with the limiting hypergraphon mean field. For this purpose, the first step is to show the convergence of agent state distributions to the mean field.

Define for any $n \in \mathbb N$ the evaluation of measurable functions $f \colon \mathbb X^n \times \mathbb I^n \to \mathbb R$ under any $n$-dimensional product measures $\bigotimes^n \boldsymbol \mu \in P(\mathbb X^n)^{\mathbb I^n}$ as
\begin{align}
    \boldsymbol \mu(f) \coloneqq \int_{\mathbb I^n} \sum_{x \in \mathbb X^n} f(x, \boldsymbol \beta) \prod_{i \in [n]} \mu^{\beta_i}(x_i) \, \mathrm d\boldsymbol \beta,
\end{align}
where $\bigotimes^n \boldsymbol \mu$ denotes the $n$-fold product of the measure $\boldsymbol \mu$, i.e. the $n$-dimensional distribution over agent states. 

Then, our first main result is the convergence of the finite-dimensional agent state marginals to the limiting deterministic mean field, given sufficient regularity of the applied policy. For this purpose, we introduce and optimize over a class $\boldsymbol \Pi_\mathrm{Lip}$ of Lipschitz-continuous policies up to at most $D_\pi$ discontinuities, i.e. $\boldsymbol \pi \in \boldsymbol \Pi_\mathrm{Lip}$ whenever $\alpha \mapsto \pi_t^\alpha$ at any time $t$ has at most $D_\pi$ discontinuities. Note however, that we could in principle approximate non-Lipschitz policies by classes of Lipschitz-continuous policies.

\begin{theorem} \label{thm:muconv}
Consider a policy $\boldsymbol \pi \in \boldsymbol \Pi_\mathrm{Lip}$ with associated mean field $\boldsymbol \mu = \Psi(\boldsymbol \pi)$. Let $(\pi^1, \ldots, \pi^N) = \mathrm{Id}_N(\boldsymbol \pi)$, $\hat \pi \in \Pi$, $t \in \mathbb T$. Under the policy tuple $(\pi^1, \ldots, \pi^{i-1}, \hat \pi, \pi^{i+1}, \ldots, \pi^N) \in \Pi^N$ and Assumption~\ref{ass:W}, we have for all finite dimensionalities $n \in \mathbb N$ and all measurable functions $f \colon \mathbb X^n \times \mathbb I^n \to \mathbb R$ uniformly bounded by fixed $M_f > 0$, that
\begin{align} \label{eq:muconv}
    &\E \left[ \left| \bigotimes^n \boldsymbol \mu^N_t(f) - \bigotimes^n \boldsymbol \mu_t(f) \right| \right] \to 0, 
\end{align}
uniformly over all possible deviations. Furthermore, the rate of convergence follows the hypergraphon convergence rate in Assumption~\ref{ass:W} up to $O(1/\sqrt N)$.  
\end{theorem}

As a special case, by considering $n=1$ and $f = \mathbbm 1_{\{x\}}$ for any $x \in \mathbb X$, we find convergence in $L_1$ of the empirical distribution of agent states $\frac 1 N \sum_{i \in [N]} \delta_{X^i_t}$ to the limiting mean field $\int_{\mathbb I} \mu^\alpha_t \, \mathrm d\alpha$.

Our second main result is the (uniform) convergence of the system for almost any agent $i \in [N]$ with deviating policy $\hat \pi \in \Pi$ to the system where the interaction with other agents is replaced by the interaction with the limiting deterministic mean field. Hence, we introduce new random variables for the single deviating agent, beginning with initial distribution $\hat X^{\frac i N}_0 \sim \mu_0$. The action variables follow the deviating policy
\begin{align}
    \hat U^{\frac i N}_t \sim \hat \pi_t(\cdot \mid \hat X^{\frac i N}_t),
\end{align}
with the state transition laws
\begin{align}
    \hat X^{\frac i N}_{t+1} \sim P_t(\cdot \mid \hat X^{\frac i N}_t, \hat U^{\frac i N}_t, \nu^{\frac i N}_t),
\end{align}
i.e. we assume that all other agents act according to their corresponding equilibrium policy $\mathrm{Id}_N(\boldsymbol \pi)$, such that the neighborhood state distributions of most agents can be replaced by the limiting term $\nu^{\frac i N}_t$ with little error in large hypergraphs.

\begin{theorem} \label{thm:xconv}
Consider a policy $\boldsymbol \pi \in \boldsymbol \Pi_\mathrm{Lip}$ with associated mean field $\boldsymbol \mu = \Psi(\boldsymbol \pi)$. Let $(\pi^1, \ldots, \pi^N) = \mathrm{Id}_N(\boldsymbol \pi)$, $\hat \pi \in \Pi$, $t \in \mathbb T$. Under the policy tuple $(\pi^1, \ldots, \pi^{i-1}, \hat \pi, \pi^{i+1}, \ldots, \pi^N) \in \Pi^N$ and Assumptions~\ref{ass:W} and \ref{ass:Lip}, for any uniformly bounded family of functions $\mathbb G$ from $\mathbb X$ to $\mathbb R$ and any $\varepsilon, p > 0$, $t \in \mathbb T$, there exists $N' \in \mathbb N$ such that for all $N > N'$
\begin{align} \label{eq:xconv}
    \sup_{g \in \mathbb G} \left| \E \left[ g(X^i_{t}) \right] - \E \left[ g(\hat X^{\frac i N}_{t}) \right] \right| < \varepsilon
\end{align}
uniformly over $\hat \pi \in \Pi, i \in \mathbb J^N$ for some $\mathbb J^N \subseteq [N]$, $|\mathbb J^N| \geq \left\lfloor (1-p) N \right\rfloor$. 

Further, for any uniformly Lipschitz, uniformly bounded family of measurable functions $\mathbb H$ from $\mathbb X \times \mathcal B(\mathbb X)$ to $\mathbb R$ and any $\varepsilon, p > 0$, $t \in \mathbb T$, there exists $N' \in \mathbb N$ such that for all $N > N'$
\begin{align} \label{eq:xmuconv}
    \sup_{h \in \mathbb H} \left| \E \left[ h(X^i_{t}, \nu^{N,i}_t) \right] - \E \left[ h(\hat X^{\frac i N}_{t}, \nu^{\frac i N}_t) \right] \right| < \varepsilon
\end{align}
uniformly over $\hat \pi \in \Pi, i \in \mathbb J^N$ for some $\mathbb J^N \subseteq [N]$ with $|\mathbb J^N| \geq \left\lfloor (1-p) N \right\rfloor$.
\end{theorem}

As a corollary, we will have good approximation of the finite hypergraph game objective through the hypergraphon mean field objective, and correspondingly the approximate Nash property of hypergraphon mean field equilibria, motivating the hypergraphon mean field game framework.

\begin{corollary} \label{coro:jconv}
Consider a policy $\boldsymbol \pi \in \boldsymbol \Pi_\mathrm{Lip}$ with associated mean field $\boldsymbol \mu = \Psi(\boldsymbol \pi)$. Let $(\pi^1, \ldots, \pi^N) = \mathrm{Id}_N(\boldsymbol \pi)$, $\hat \pi \in \Pi$, $t \in \mathbb T$. Under the policy tuple $(\pi^1, \ldots, \pi^{i-1}, \hat \pi, \pi^{i+1}, \ldots, \pi^N) \in \Pi^N$ and Assumptions~\ref{ass:W} and \ref{ass:Lip}, there exists $N' \in \mathbb N$ such that for all $N > N'$ we have
\begin{align} \label{eq:Jconv}
    \left| J_i^N(\pi^1, \ldots, \pi^{i-1}, \hat \pi, \pi^{i+1}, \ldots, \pi^N) - J^{\boldsymbol \mu}_{\frac i N}(\hat \pi) \right| < \varepsilon
\end{align}
uniformly over $\hat \pi \in \Pi, i \in \mathbb J^N$ for some $\mathbb J^N \subseteq [N]$ with $|\mathbb J^N| \geq \left\lfloor (1-p) N \right\rfloor$.
\end{corollary}

\begin{corollary} \label{coro:approxnash}
Consider an HMFE $(\boldsymbol \pi, \boldsymbol \mu) \in \boldsymbol \Pi_\mathrm{Lip} \times \mathbb M$. Under Assumptions~\ref{ass:W} and \ref{ass:Lip}, for any $\varepsilon, p > 0$ there exists $N'$ such that for all $N > N'$, the policy $(\pi^1, \ldots, \pi^N) = \mathrm{Id}_N(\boldsymbol \pi)$ is an $(\varepsilon, p)$-Nash equilibrium.
\end{corollary}

Therefore, we find that a solution of the mean field system is a good equilibrium solution of sufficiently large finite hypergraph games. 

The assumption of a class $\boldsymbol \Pi_\mathrm{Lip}$ of Lipschitz continuous policies up to finitely many discontinuities may seem restrictive. However -- similar to \citet[Theorem~5]{cui2022learning} -- we may discretize and partition $\mathbb I$ in order to solve hypergraphon mean field games to an arbitrary degree of exactness, preserving the good approximation properties on large hypergraph games.

\section{\label{sec:exp}Numerical Experiments}
In this section, we shall introduce an exemplary numerical problem of rumor spreading, and show associated numerical solutions to demonstrate the hypergraphon mean field framework, verifying the theoretical results.

In order to learn an HMFE in our model, we shall adopt the well-founded discretization method proposed in \citet{cui2022learning} analogous to the technique used in the proof of Theorem~\ref{thm:existence} to convert the graphon mean field game into a classical mean field game, and thereby allow application of any existing mean field game algorithms such as fixed point iteration to solve for an equilibrium. In other words, we will split $\mathbb I$ into subintervals $I_1^N, \ldots, I_N^N$, for each of which we will pick a representing $\alpha \in I_i^N$. This $\alpha$ together with an agent's original state in $\mathbb X$ will constitute the new state. In Appendix~\ref{app:exp}, we perform additional experiments for another numerical problem of epidemics control, where existing algorithms fail, pointing out potential future work.

\subsection{Hypergraphons}
In our experiments, we shall sample finite hypergraphs directly from given limiting hypergraphons, which should ensure that we obtain hypergraph sequences that fulfill Assumption~\ref{ass:W} analogous to the standard graphon case at rate $O(\frac{1}{\sqrt{\log N}})$, see \citet[Lemma~10.16]{lovasz2012large}. To sample a $k$-uniform hypergraph with $N$ nodes from a $k$-uniform hypergraphon $W$, we sample $|r_<[k]|$ uniformly distributed values from the unit interval $\{ \alpha_{\mathbf j} \colon \alpha_{\mathbf j} \sim \mathrm{Unif}([0,1])\}_{\mathbf j \in r_<[k]}$. Then, we add any hyperedge $B \subseteq [N]$ with probability $W(\alpha_{r_<(B)})$.

\begin{figure}
    \centering
    \includegraphics[width=\linewidth]{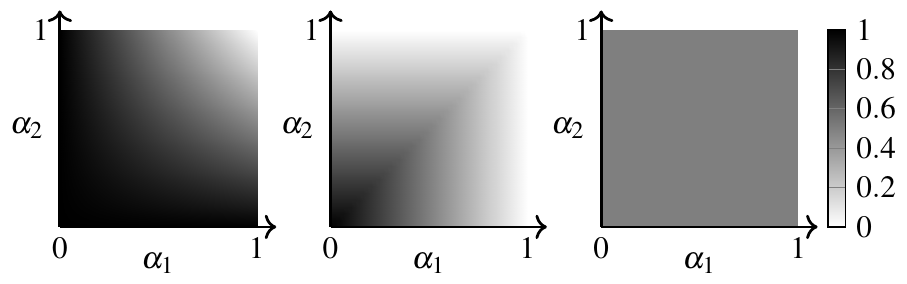}
    \caption{Visualization of example graphons in the $2$-dimensional case. Left: Uniform attachment graphon; Middle: Ranked attachment graphon; Right: $0.5$-ER graphon.}
    \label{fig:examplegraphons}
\end{figure}

For the sake of illustration, unless otherwise noted, we will in the following consider two-layer hypergraphons, where the first layer is a $2$-uniform hypergraph (standard graph), while the second layer shall be a $3$-uniform hypergraph. For the first layer, we consider the uniform attachment graphon 
\begin{align*}
    W_\mathrm{unif}(\alpha_1, \alpha_2) = 1 - \max(\alpha_1, \alpha_2),
\end{align*}
the ranked attachment graphon 
\begin{align*}
    W_\mathrm{rank}(\alpha_1, \alpha_2) = 1 - \alpha_1 \alpha_2
\end{align*}
and the flat (or $p$-ER) random graphon 
\begin{align*}
    W_\mathrm{flat} \coloneqq p=0.5.
\end{align*}
In particular, the uniform attachment graphon is the limit of a random graph sequence where we iteratively add a new node $N$ and then connect all unconnected nodes with probability $\frac 1 N$. Similarly, for the ranked attachment graphon, at each iteration $n$ we first add a new ($n$-th) node. Before adding the node, the nodes $1, \ldots, n-1$ exist from prior iterations. The new node $n$ is connected to all previous nodes $i=1,\ldots,n-1$ with probability $1 - \frac i n$. Then, all other nodes that are not yet connected with each other will connect with probability $\frac 2 n$. See also \citet[Chapter~11]{lovasz2012large} and Fig.~\ref{fig:examplegraphons}. 

For the second, $3$-uniform layer, we similarly consider the hypergraphon resulting from converting all triangles in a standard $p$-ER graph into hyperedges \cite{zhao2015hypergraph}
\begin{align*}
    \hat W_\mathrm{ind}(\boldsymbol \alpha) \coloneqq \mathbbm 1_{\mathbb I^3 \times [0, p]^3}(\alpha), 
\end{align*}
as well as the uniform attachment hypergraphon 
\begin{align*}
    \hat W_\mathrm{unif}(\boldsymbol \alpha) = 1 - \max(\alpha_1, \alpha_2, \alpha_3)
\end{align*} and its inverted version 
\begin{align*}
    \hat W_\mathrm{inv-unif}(\boldsymbol \alpha) = 1 - \max(1-\alpha_1, 1-\alpha_2, 1-\alpha_3)
\end{align*} 
resulting from a similar construction as in the standard case.

\begin{figure*}
    \centering
    \includegraphics[width=\linewidth]{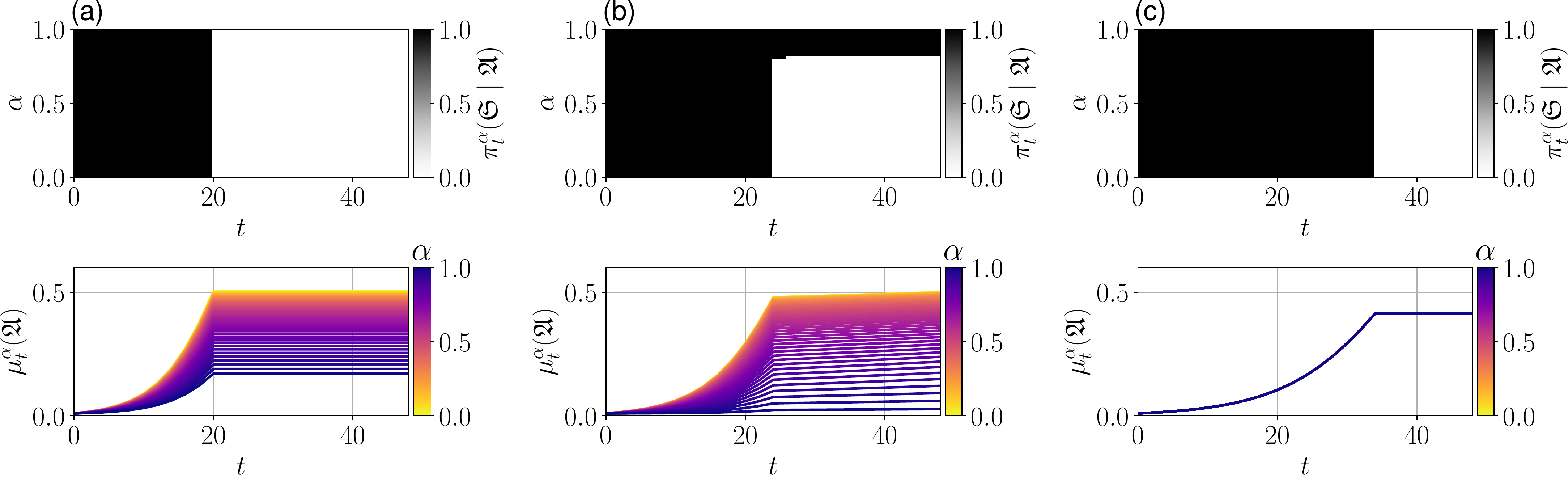}
    \caption{Analysis of equilibrium behavior for the Rumor problem. Top: The threshold policy allows spreading of rumors. It can be seen that agents spread the rumor up until a point in time where too many other agents know of the rumor. As expected, agents are more likely to hear of the rumor if they have more neighbors. (a): $(W_\mathrm{rank}, \hat W_\mathrm{unif})$; (b): $(W_\mathrm{unif}, \hat W_\mathrm{unif})$; (c): $(W_\mathrm{er}, \hat W_\mathrm{ind})$.}
    \label{fig:rumor}
\end{figure*}

\subsection{Rumor spreading dynamics}
In this section, we will describe some simple social dynamics and epidemics problems to illustrate potential applications of hypergraphon mean field games. Here, each layer could model different types of interpersonal relationships. In our particular example of $2$-uniform and $3$-uniform layers, the latter can model small cliques of friends, while the former could model general acquaintanceship. We do note that social networks are typically more sparse, possessing significantly less edges than on the order of $O(N^2)$. However, our model is a first step towards rigorous limiting hypergraph models and in the future could be extended by using other graph limit theories such as $L^p$ graphons \cite{borgs2018p, borgs2019p, fabian2022learning} by extending their theory towards hypergraphons. We further imagine that similar approaches could be used e.g. in economics \cite{carmona2020applications} or engineering applications \cite{djehiche2017mean}.

In the classical Maki-Thompson model \cite{maki1973mathematical, junior2020maki}, spread of rumors is modelled via three node states: ignorant, spreader and stifler. Ignorants are unaware of the rumor, while spreaders attempt to spread the rumor. When spreaders attempt to spread to nodes that are already aware of the rumor too often, they stop spreading and become a stifler. In this work, instead of a priori assuming the above behavior, we will give agents an intrinsic motivation to spread or stifle rumors, giving rise to the Rumor problem. We shall consider ignorant ($\mathfrak I$) and aware ($\mathfrak A$) nodes. The behavior of aware nodes is then motivated by the gain and loss of social standing resulting from spreading rumors to ignorant and aware nodes respectively. The possible actions $\mathbb U \coloneqq \{\bar{\mathfrak S}, \mathfrak S\}$ of nodes are to actively spread the rumor ($\mathfrak S$) or to refrain from doing so ($\bar{\mathfrak S}$). The probability of an ignorant node becoming aware of the rumor at any decision epoch is then simply given by a linear combination of all layer neighborhood densities of aware, spreading nodes. 

Since transition dynamics will depend on the spreading actions of neighbors, following Remark~\ref{remark:actiondep} we define instead the extended state space $\mathbb X = \{\mathfrak I, \mathfrak A\} \cup (\{\mathfrak I, \mathfrak A\} \times \mathbb U)$. We then assume the dynamics are given at all times by
\begin{align*}
    P((x, u) \mid x, u, \cdot) &= 1, \quad P(\mathfrak A \mid (\mathfrak A, \cdot), \cdot, \boldsymbol \nu) = 1, \\
    P(\mathfrak A \mid (\mathfrak I, \cdot), \cdot, \boldsymbol \nu) &= 1 - P(\mathfrak I \mid (\mathfrak I, \cdot), \cdot, \boldsymbol \nu) \\
    &= \min \left( 1, \sum_{d \in [D]} \tau_d \boldsymbol \nu_d \left( \sum_{i \in [k_d]} \mathbbm 1_{\{(\mathfrak A, \mathfrak S)\}}(\cdot_i) \right) \right)
\end{align*}
for all $x \in \{\mathfrak I, \mathfrak A\}$, $u \in \mathbb U$ and similarly the rewards
\begin{multline*}
    R((\mathfrak A, \mathfrak S), \cdot, \boldsymbol \nu) \\= \sum_{d \in [D]} \boldsymbol \nu_d \left( \sum_{i \in [k_d]} r_d \mathbbm 1_{\{\mathfrak I\} \times \mathbb U}(\cdot_i) - c_d \mathbbm 1_{\{\mathfrak A\} \times \mathbb U}(\cdot_i) \right)
\end{multline*}
with $R \equiv 0$ otherwise. In other words, any aware and spreading agent obtains a reward in each layer that is proportional to the probability of a neighbor of any hyperedge sampled uniformly-at-random out of all connected hyperedges to be ignorant. In our experiments, we use $\tau_1 = 0.3$, $\tau_2 = 0.5$, $r_d = 0.5$, $c_d = 0.8$, $\mu_0(\mathfrak A) = 0.01$ and $\mathbb T = \{0, 1, \ldots, 49\}$.

\subsection{Numerical Results}
In our experiments, we restrict ourselves to finite time horizons with $\gamma=1$, $50$ discretization points, and use backwards induction with exact forward propagation to compute exact solutions. Note that simple fixed point iteration by repeatedly computing an arbitrary optimal deterministic policy and its corresponding mean field converges to an equilibrium in the Rumor problem. In general however, fixed point iteration (as well as more advanced state-of-the-art techniques) may fail to converge, see e.g. the SIS problem in Appendix~\ref{app:exp}.

\begin{figure}[t]
    \centering
    \includegraphics[width=\linewidth]{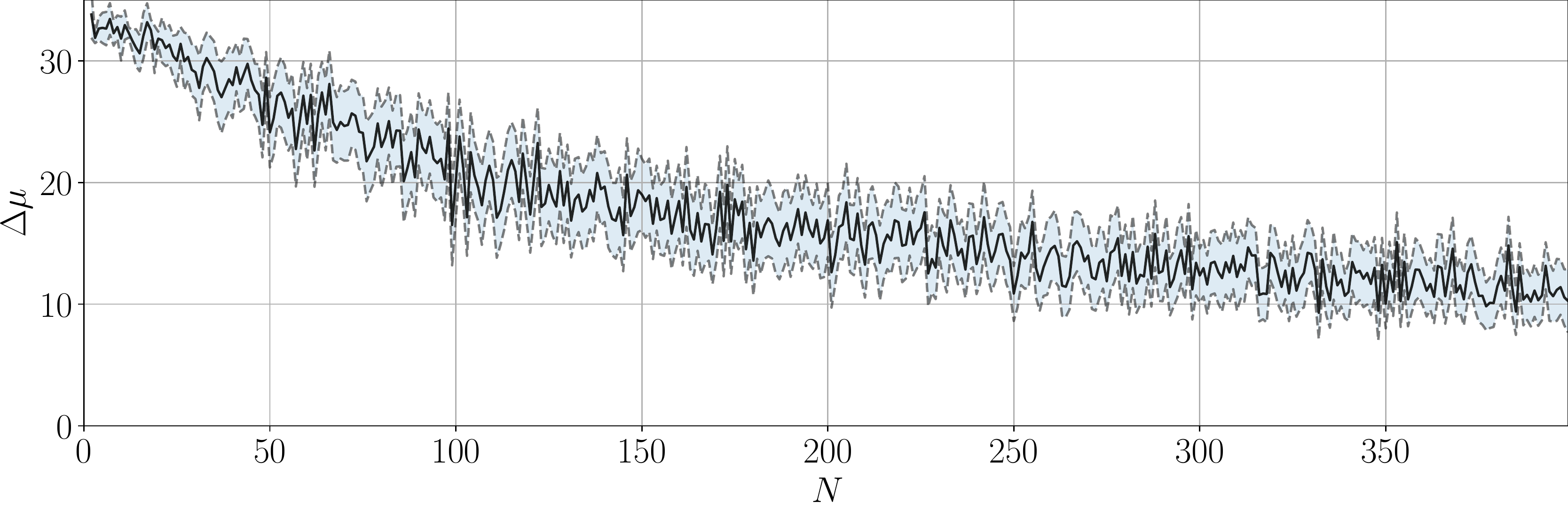}
    \caption{Comparison between the fraction of aware nodes in the finite and mean field system under the equilibrium policy for $(W_\mathrm{rank}, \hat W_\mathrm{inv-unif})$ from Fig.~\ref{fig:rumormore}(a) in Appendix~\ref{app:exp}, averaged over $50$ stochastic simulations. The shaded region depicts the $95\%$ confidence interval at each $N$. It can be seen that the state distributions are increasingly well approximated by the mean field.}
    \label{fig:results_finite}
\end{figure}

\begin{figure}[b]
    \centering
    \includegraphics[width=\linewidth]{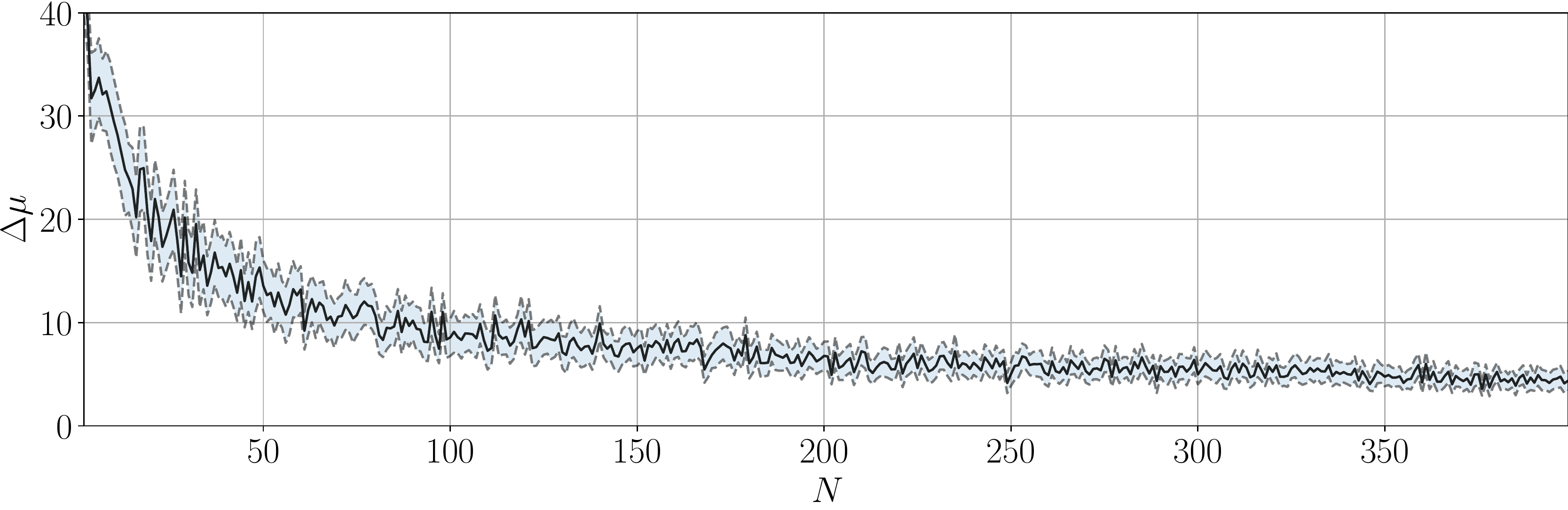}
    \caption{Comparison between the fraction of aware nodes in the finite and mean field system under the equilibrium policy for $(W_\mathrm{rank}, \hat W_\mathrm{inv-unif})$ as in Fig.~\ref{fig:results_finite}, but with higher initial awareness. It can be seen that convergence is much faster, since the effect of random sparse initialization is avoided.}
    \label{fig:results_finite2}
\end{figure}

\begin{figure}[bt]
    \centering
    \includegraphics[width=\linewidth]{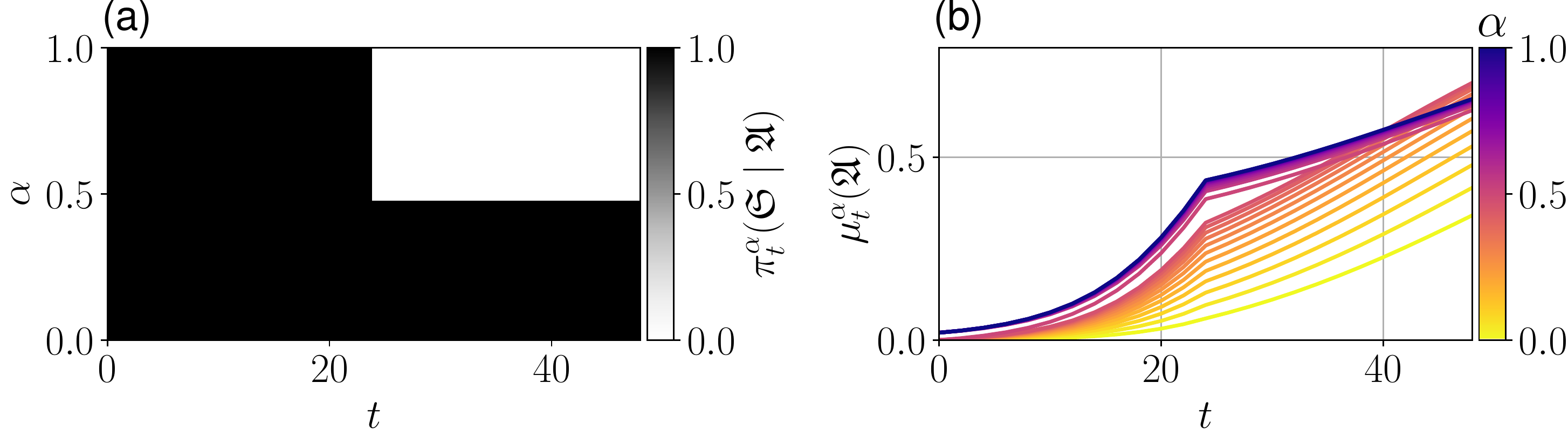}
    \caption{Analysis of equilibrium behavior for the Rumor problem with two-layer $3$-uniform hypergraphs $(\hat W_\mathrm{block}, \hat W_\mathrm{unif})$. We observe that the rumor originates in nodes with $\alpha > 0.5$, but nodes with $\alpha \leq 0.5$ eventually catch up due to their increased connectivity. (a): The equilibrium threshold policy. (b): Mean field for each $\alpha$.}
    \label{fig:multi}
\end{figure}

In Fig.~\ref{fig:rumor}, we can observe that the behavior for the Rumor problem is as expected. At the equilibrium, agents will continue to spread rumors until the number of aware agents reaches a critical point at which the penalty for spreading to aware agents is larger than the reward for spreading to ignorant agents. The agents with higher connectivity are more likely to be aware of the rumor. Particularly in the uniform attachment hypergraphon case, the threshold is reached at different times, since the neighborhoods of different $\alpha$ reach awareness at different rates depending on their connectivity. Here, a number of nodes with very low degrees will continue spreading the rumors. In Appendix~\ref{app:exp}, we show additional results for inverted $3$-uniform hypergraphons, which give similar results to the ones seen here. Furthermore, as can be seen in Fig.~\ref{fig:results_finite}, the $L_1$ error between the empirical distribution and the limiting mean field system (as vectors over time)
\begin{align}
    \Delta \mu = \mathbb E \left[ \sum_{x \in \mathbb X} \sum_{t \in \mathbb T} \left| \frac 1 N \sum_{i \in [N]} \delta_{X^i_t}(x) - \int_{\mathbb I} \mu^\alpha_t(x) \, \mathrm d\alpha \right| \right]
\end{align}
goes to zero as the number of agents increases, showing that the finite hypergraph game is well approximated by the hypergraphon mean field game for sufficiently large systems, though the error remains somewhat large due to the high variance from our sparse initialization $\mu_0(\mathfrak I) = 0.01$. Here, we estimated the error $\Delta \mu$ for each $N$ over $50$ realizations. Due to the $O(N^2)$ complexity of simulation and computational constraints, our experiments remain limited to the demonstrated number of agents. We repeat the experiment in Fig.~\ref{fig:results_finite2} with a more dense initialization $\mu_0(\mathfrak A) = 0.1$ to reduce the aforementioned high contribution of variance from random initializations. Here, we observe that the resulting convergence is significantly faster.

Lastly, in Fig.~\ref{fig:multi} we demonstrate some interesting non-linear behavior for a two-layer setting where both layers consist of $3$-uniform hypergraphs. Here, for the first layer we use the block hypergraphon
\begin{align*}
    \hat W_\mathrm{block}(\boldsymbol \alpha) \coloneqq \mathbbm 1_{\mathbb [0, 0.5]^3 \times [0, p]^3}(\alpha) + \mathbbm 1_{\mathbb (0.5, 1]^3 \times [0, p]^3}(\alpha), 
\end{align*}
for $p=0.5$, while for the second layer we again use the inverted uniform attachment hypergraphon. In other words, we have a structure of two blocks on the first layer, while the second layer is more globally connected. Furthermore, we will initialize the rumor in the second block where $\alpha > 0.5$, i.e. $\mu_0^\alpha(\mathfrak A) = \mathbbm 1_{(0.5, 1]}(\alpha)$. As we can see in Fig.~\ref{fig:multi}, in the beginning the rumor spreads in the second block $\alpha > 0.5$ where it originated from. After a while however, the rumor begins to spread faster in the first block $\alpha \leq 0.5$, since nodes with low $\alpha$ are significantly more interconnected on the second layer. 

Overall, we can see that multi-layer hypergraphon mean field games allow for more complex behavior and modelling of connections than a single-layer graphon approach.

\section{\label{sec:concl}Conclusion}
In this work, we introduced a model for dynamical systems on hypergraphs that can describe agents with weak interaction via the graph structure. The model allows for a rigorous and simple mean field description that has a complexity independent of the number of agents. We verify our approach both theoretically and empirically on a rumor spreading example. By introducing game-theoretical ideas, we thus obtain a framework for solving otherwise intractable large-scale games on hypergraphs in a tractable manner. 

We hope our work forms the basis for several future works, e.g., extensions to directed or weighted hypergraphs in order to generalize to arbitrary network motifs \cite{cui2022motif}, adaptive networks \cite{garbe2022flip}, cooperative control or consideration of edge states in addition to the vertex states we have considered in this work. Furthermore, it may be of interest to consider graph models with more adjustable clustering parameters. An extension of our rumor model and theory to continuous-time models could be fruitful. Finally, so far our work remains restricted to dense graphs and deterministic limiting graphons, while in practice this is not always the case (e.g. preferential attachment graphs \cite{borgs2011limits}). Here, $L^p$ graphons \cite{borgs2018p, borgs2019p, fabian2022learning} could provide a description for less dense cases, which are of great practical interest. We also hope that our work inspires future applications in inherently (hyper-)graphical scenarios.

\begin{acknowledgments}
This work has been funded by the LOEWE initiative (Hesse, Germany) within the emergenCITY center. The authors also acknowledge support by the German Research Foundation (DFG) via the Collaborative Research Center (CRC) 1053 – MAKI. WRK received no specific grant for this research from any funding agency in the public, commercial, or not-for-profit sectors. The authors acknowledge the Lichtenberg high performance computing cluster of the TU Darmstadt for providing computational facilities for the calculations of this research.
\end{acknowledgments}

\section*{Data Availability Statement}
Data sharing is not applicable to this article as no new data were created or analyzed in this study.

\appendix

\section{Proofs}
\subsection{Proof of Theorem~\ref{thm:existence}}
\begin{proof}
Under our assumptions, we can verify \citet[Assumption~1]{saldi2018markov} for the equivalent standard mean field game given by \eqref{eq:inducMDP} as in \citet{cui2022learning}. By \citet[Theorem~3.3]{saldi2018markov} there exists a mean field equilibrium $(\tilde \pi, \tilde \mu)$ for \eqref{eq:inducMDP}. The policy $\tilde \mu$ is $\alpha$-a.e. optimal under the mean field $\tilde \mu$ by \citet[Theorem 3.6]{saldi2018markov}. For all other $\alpha$, there trivially exists an optimal action, i.e. we can change $\tilde \pi$ such that it is optimal for all $\alpha$. Since the change is on a null set of $\mathbb I$, $(\tilde \pi, \tilde \mu)$ remains a mean field equilibrium. Define the hypergraphon mean field policy $\boldsymbol \pi$ by $\pi_t^\alpha(u \mid x) = \tilde \pi_t(u \mid x, \alpha)$, then $\boldsymbol \pi$ is optimal under the hypergraphon mean field $\boldsymbol \mu$ where $\boldsymbol \mu = \Psi(\boldsymbol \pi)$, since $\mu_t^\alpha = \tilde \mu_t(\cdot, \alpha)$ for almost every $\alpha$. Finally, both $\boldsymbol \pi$ and $\boldsymbol \mu$ are measurable. Therefore, we have proven existence of the HMFE $(\boldsymbol \pi, \boldsymbol \mu)$.
\end{proof}

\subsection{Proof of Theorem~\ref{thm:muconv}}
In this section, we provide the full proof of Theorem~\ref{thm:muconv}. In contrast to prior work such as \citet{cui2022learning}, we (i) extend existing mean field convergence results to $n$-fold products of the state distributions; and (ii) replace the state distributions by their symmetrized version, in order to obtain convergence results under the generalized cut norm \eqref{eq:cutconv}. Propagating these changes forward, the rest of the proof is (somewhat) readily generalized and given in the following.

To begin, we introduce some notation to improve readability. Define the $D$-dimensional neighborhood mean fields $\nu^{\alpha, \boldsymbol \mu}_{W}$ with $d$-th component
\begin{align*}
    \nu^{\alpha, \boldsymbol \mu}_{W,d} \coloneqq \int_{\mathbb I^{r_<[k_d] \setminus \{1\}}} W_d(\alpha, \boldsymbol \beta) \prod_{j=1}^{k_d-1} \mu^{\beta_j}(\cdot) \, \mathrm d\boldsymbol \beta
\end{align*}
for all $\boldsymbol \mu \in \mathcal P(\mathbb X)^{\mathbb I}$, $W \coloneqq (W_1,\ldots,W_D) \in \bigtimes_{d=1}^D \mathbb W_{k_d}$ as well as the transition operator $P^{t, \boldsymbol \pi, \boldsymbol \mu}_{W} \colon \mathcal P(\mathbb X)^{\mathbb I} \to \mathcal P(\mathbb X)^{\mathbb I}$ such that
\begin{multline*}
    \left( \boldsymbol \mu' P^{t, \boldsymbol \pi, \boldsymbol \mu}_{W} \right)^\alpha = \sum_{x \in \mathbb X} \mu'^\alpha(x) \sum_{u \in \mathbb U} \pi^\alpha(u \mid x) P_t \left(\cdot \innermid x, u, \nu^{\alpha, \boldsymbol \mu}_{W} \right)
\end{multline*}
for all $\mu' \in \mathcal P(\mathbb X)^{\mathbb I}$, $\boldsymbol \pi \in \mathcal P(\mathbb U)^{\mathbb X}$, such that e.g.
\begin{align*}
    \boldsymbol \mu_{t+1} = \boldsymbol \mu_t P^{t, \boldsymbol \pi_t, \boldsymbol \mu_t}_{W}.
\end{align*}

\begin{proof}
In the following, consider arbitrary measurable functions $f \colon \mathbb X \times \mathbb I \to [-M_f, M_f]$, $M_f > 0$ and the telescoping sum
\begin{align*}
    &\E \left[ \left| \bigotimes^n \boldsymbol \mu^N_t(f) - \bigotimes^n \boldsymbol \mu_t(f) \right| \right] \\
    &\quad \leq \sum_{i=0}^{n-1} \E \left[ \left| \bigotimes^{n-i} \boldsymbol \mu^N_t \otimes \bigotimes^{i} \boldsymbol \mu_t(f) - \bigotimes^{n-i-1} \boldsymbol \mu^N_t \otimes \bigotimes^{i+1} \boldsymbol \mu_t (f) \right| \right] \\
    &\quad = \sum_{i=0}^{n-1} \E \left[ \left| \bigotimes^{n-i-1} \boldsymbol \mu^N_t \otimes \left( \boldsymbol \mu^N_t - \boldsymbol \mu_t \right) \otimes \bigotimes^{i} \boldsymbol \mu_t(f) (f) \right| \right] \\
    &\quad = \sum_{i=0}^{n-1} \E \left[ \left| \int_{\mathbb I} \sum_{x_i \in \mathbb X} \int_{\mathbb I^{[n] \setminus \{i\}}} \sum_{\tilde x \in \mathbb X^{[n] \setminus \{i\}}} f(x, (\alpha, \boldsymbol \beta))
    \right.\right.\\&\hspace{2.5cm} \left.\left. 
    \cdot \prod_{j=1}^{n-i-1} \mu^{N,\beta_j}(\tilde x_j) \prod_{j=n-i+1}^{n} \mu^{\beta_j}(\tilde x_j) \, \mathrm d\boldsymbol \beta 
    \right.\right.\\&\hspace{4cm} \left.\left. 
    \cdot \left[ \mu^{N,\alpha}(x_i) - \mu^{\alpha}(x_i) \right] \, \mathrm d\alpha \right| \right]  \\
    &\quad = \sum_{i=0}^{n-1} \E \left[ \left| \int_{\mathbb I} \sum_{x_i \in \mathbb X} g(x_i, \alpha) \left[ \mu^{N,\alpha}(x_i) - \mu^{\alpha}(x_i) \right] \, \mathrm d\alpha \right| \right] 
\end{align*}
where we defined $g \colon \mathbb X \times \mathbb I \to [-M_f, M_f]$ as
\begin{multline*}
    g(x, \alpha) \coloneqq \int_{\mathbb I^{[n] \setminus \{i\}}} \sum_{x_{-i} \in \mathbb X^{[n] \setminus \{i\}}} f((x, x_{-i}), (\alpha, \boldsymbol \beta)) \\ \cdot \prod_{j=1}^{n-i-1} \mu^{N,\beta_j}(x_j) \prod_{j=n-i+1}^{n} \mu^{\beta_j}(x_j) \, \mathrm d\boldsymbol \beta.
\end{multline*}

Since $g$ is a measurable function bounded by $M_f$, due to the prequel it suffices at any time $t \in \mathbb T$ to prove \eqref{eq:muconv} for $n=1$, which will imply the statement for all $n \in \mathbb N$. 

The proof is by induction over $t$ for $n=1$. At $t=0$, 
\begin{align*}
    &\E \left[ \left| \boldsymbol \mu^N_{0}(f) - \boldsymbol \mu_{0}(f) \right| \right] \\
    &\quad = \E \left[ \left| \int_{\mathbb I} \sum_{x \in \mathbb X} \mu^{N, \alpha}_{0}(x) \, f(x, \alpha) - \sum_{x \in \mathbb X} \mu^\alpha_0(x) \, f(x, \alpha) \, \mathrm d\alpha \right| \right] \\
    &\quad = \E \left[ \left| \sum_{i \in [N]} \left( \int_{I_i^N} f(X^i_{0}, \alpha) \, \mathrm d\alpha - \E \left[ \int_{I_i^N} f(X^i_{0}, \alpha) \, \mathrm d\alpha \right] \right) \right| \right] \\
    &\quad \leq \left( \V \left[ \sum_{i \in [N]} \int_{I_i^N} f(X^i_{0}, \alpha) \, \mathrm d\alpha \right] \right)^{\frac 1 2} \\
    &\quad = \left( \sum_{i \in [N]} \V \left[ \int_{I_i^N} f(X^i_{0}, \alpha) \, \mathrm d\alpha \right] \right)^{\frac 1 2} \leq \frac{4 M_f}{\sqrt N}
\end{align*}
by i.i.d. $X^i_0 \sim \mu_0 = \mu^\alpha_0$ and $\V \left[ \int_{I_i^N} f(X^i_{0}, \alpha) \, \mathrm d\alpha \right] \leq \left( \frac{4 M_f}{N} \right)^2$.

Assume that \eqref{eq:muconv} holds at $t \in \mathbb T$. Then at time $t+1$ we have
\begin{align*}
    &\E \left[ \left| \boldsymbol \mu^N_{t+1}(f) - \boldsymbol \mu_{t+1}(f) \right| \right] \\
    &\quad \leq \E \left[ \left| \boldsymbol \mu^N_{t+1}(f) - \boldsymbol \mu^N_t P^{t, \boldsymbol \pi^N_t, \boldsymbol \mu^N_t}_{W^N}(f) \right| \right] \\
    &\qquad + \E \left[ \left| \boldsymbol \mu^N_t P^{t, \boldsymbol \pi^N_t, \boldsymbol \mu^N_t}_{W^N}(f) - \boldsymbol \mu^N_t P^{t, \boldsymbol \pi^N_t, \boldsymbol \mu^N_t}_{W}(f) \right| \right] \\
    &\qquad + \E \left[ \left| \boldsymbol \mu^N_t P^{t, \boldsymbol \pi^N_t, \boldsymbol \mu^N_t}_{W}(f) - \boldsymbol \mu^N_t P^{t, \boldsymbol \pi_t, \boldsymbol \mu^N_t}_{W}(f) \right| \right] \\
    &\qquad + \E \left[ \left| \boldsymbol \mu^N_t P^{t, \boldsymbol \pi_t, \boldsymbol \mu^N_t}_{W}(f) - \boldsymbol \mu^N_t P^{t, \boldsymbol \pi_t, \boldsymbol \mu_t}_{W}(f) \right| \right] \\
    &\qquad + \E \left[ \left| \boldsymbol \mu^N_t P^{t, \boldsymbol \pi_t, \boldsymbol \mu_t}_{W}(f) - \boldsymbol \mu_{t+1}(f) \right| \right]
\end{align*}
and in the following we will analyze each term.

For the first term, observe first that by definition,
\begin{align*}
    &\int_{\mathbb I^{r_<[k_d] \setminus \{1\}}} W^N_d(\alpha, \boldsymbol \beta) \prod_{j=1}^{k_d-1} \mu^{N,\beta_j}_t(\cdot) \, \mathrm d\boldsymbol \beta \\
    &\quad = \frac 1 {N^{k_d-1}} \sum_{\mathbf m \in [N]^{k_d-1}} \mathbbm 1_{E^d[H_N]}(\mathbf m \cup i) \delta_{\bigtimes_{j\neq i} X^{m_j}_t}
\end{align*}
and therefore
\begin{align*}
    &\boldsymbol \mu^N_t P^{t, \boldsymbol \pi^N_t, \boldsymbol \mu^N_t}_{W^N}(f) = \E \left[ \int_{I_i^N} f(X^i_{t+1}, \alpha) \, \mathrm d\alpha \innermid \mathbf X_t \right]
\end{align*}
such that we again obtain
\begin{align*}
    &\E \left[ \left| \boldsymbol \mu^N_{t+1}(f) - \boldsymbol \mu^N_t P^{t, \boldsymbol \pi^N_t, \boldsymbol \mu^N_t}_{W^N}(f) \right| \right] \\
    &\quad = \E \left[ \left| \sum_{i \in [N]} \left( g(X^i_{t+1}) - \E \left[ g(X^i_{t+1}) \innermid \mathbf X_t \right] \right) \right| \right] \\
    &\quad \leq \left( \E \left[ \left( \sum_{i \in [N]} \left( g(X^i_{t+1}) - \E \left[ g(X^i_{t+1}) \innermid \mathbf X_t \right] \right) \right)^2 \right] \right)^{\frac 1 2} \\
    &\quad = \left( \sum_{i \in [N]} \E \left[ \left( g(X^i_{t+1}) - \E \left[ g(X^i_{t+1}) \innermid \mathbf X_t \right] \right)^2 \right] \right)^{\frac 1 2} \leq \frac{4 M_f}{\sqrt N}
\end{align*}
where $g(x) \coloneqq \int_{I_i^N} f(x, \alpha) \, \mathrm d\alpha$, $|g| \leq \frac {M_f} N$, by using the law of total expectation and conditional independence of $\{ X^i_{t+1} \}_{i \in [N]}$ given $\mathbf X_t \coloneqq \{ X^i_{t} \}_{i \in [N]}$.

For the second term, first note that we can replace the distributional terms by their symmetrized version: For any $k \in \mathbb N$, any $W \in \operatorname{Sym}^\mathrm{ind}_<[k]$ and any step empirical measure or mean field $\boldsymbol \mu \in \mathcal P(\mathbb X)^{\mathbb I}$, we have by symmetry that the associated neighborhood probabilities are invariant to all permutations $\sigma \in \operatorname{Sym}([k-1])$ of states $\mathbb X^{k-1}$, i.e. for any $x \in \mathbb X^{k-1}$, $\alpha \in \mathbb I$
\begin{align*}
    &\int_{\mathbb I^{r_<[k] \setminus \{1\}}} W(\alpha, \beta) \prod_{i=1}^{k-1} \mu^{\beta_i}(x_i)) \, \mathrm d\beta \\
    &\quad = \frac 1 {(k-1)!} \sum_{\sigma \in \operatorname{Sym}([k-1])} \int_{\mathbb I^{r_<[k] \setminus \{1\}}} W(\alpha, \beta) \prod_{i=1}^{k-1} \mu^{\beta_i}(x_{\sigma(i)})) \, \mathrm d\beta \\
    &\quad = \int_{\mathbb I^{r_<[k] \setminus \{1\}}} W(\alpha, \beta) \underbrace{\frac 1 {(k-1)!} \sum_{\sigma \in \operatorname{Sym}([k-1])} \prod_{i=1}^{k-1} \mu^{\beta_i}(x_{\sigma(i)}))}_{u_1 \in \operatorname{Sym}^\mathrm{ind}_\leq[k-1]} \, \mathrm d\beta
\end{align*}
and hence Assumption~\ref{ass:W} implies that
\begin{align*}
    &\int_{\mathbb I} \left| \int_{\mathbb I^{r_<[k] \setminus \{1\}}} W(\alpha, \beta) u_1(\beta_{[k] \setminus \{1\}]}) \, \mathrm d\beta \right| \, \mathrm d\alpha \\
    &\quad \leq \int_{\mathbb I^{[k]}} \left| \int_{\mathbb I^{r_<[k] \setminus [k]}} W(\alpha, \beta) u_1(\alpha_{[k] \setminus \{1\}]}) \, \mathrm d\beta \right| \, \mathrm d\alpha \\
    &\quad \leq \sup_{\substack{u_1 \colon \mathbb I^{r[k-1]} \to \mathbb I, \\u_1 \in \operatorname{Sym}^\mathrm{ind}_\leq[k-1]}} \int_{\mathbb I^{[k]}} \left| \int_{\mathbb I^{r_<[k] \setminus [k]}} W(\alpha, \beta) \, \mathrm d\beta u_1(\alpha_{[k] \setminus \{1\}]}) \right| \, \mathrm d\alpha \\
    &\quad = \sup_{\substack{u_1 \colon \mathbb I^{r[k-1]} \to \mathbb I, \\u_1 \in \operatorname{Sym}^\mathrm{ind}_\leq[k-1]}} \int_{\mathbb I^{[k]}} \int_{\mathbb I^{r_<[k] \setminus [k]}} W(\alpha, \beta) \, \mathrm d\beta u_1(\alpha_{[k] \setminus \{1\}]}) \, \mathrm d\alpha \\
    &\quad \leq \lVert W \rVert_{\square^{k-1}} \to 0
\end{align*}
for any $x \in \mathbb X^{k-1}$, $\alpha \in \mathbb I$ by letting $u_1(\alpha_{r_<([k] \setminus \{i\}})) \coloneqq \frac 1 {(k-1)!} \sum_{\sigma \in \operatorname{Sym}([k-1])} \prod_{i=1}^{k-1} \mu^{\beta_i}(x_{\sigma(i)}))$ in \eqref{eq:cutconv}. Therefore,
\begin{align*}
    &\E \left[ \left| \boldsymbol \mu^N_t P^{t, \boldsymbol \pi^N_t, \boldsymbol \mu^N_t}_{W^N}(f) - \boldsymbol \mu^N_t P^{t, \boldsymbol \pi^Nt, \boldsymbol \mu^N_t}_{W}(f) \right| \right] \\
    &\quad = \E \left[ \left| \int_{\mathbb I} \sum_{x \in \mathbb X} \mu^{N,\alpha}_t(x) \sum_{u \in \mathbb U} \pi^{N,\alpha}_t(u \mid x) \sum_{x' \in \mathbb X} f(x', \alpha) 
    \right.\right.\\&\quad \left.\left. 
    \cdot \left[ P_t \left(x' \mid x, u, \nu^{\alpha, \boldsymbol \mu^N_t}_{W^N} \right) - P_t \left(x' \mid x, u, \nu^{\alpha, \boldsymbol \mu^N_t}_{W} \right) \right] \, \mathrm d\alpha \right| \right] \\
    &\quad \leq |\mathbb X| M_f L_P\E \left[ \int_{\mathbb I} \left\Vert \nu^{\alpha, \boldsymbol \mu^N_t}_{W^N} - \nu^{\alpha, \boldsymbol \mu^N_t}_{W} \right\Vert \, \mathrm d\alpha \right] \\
    &\quad \leq |\mathbb X| M_f L_P\E \left[ \int_{\mathbb I} \sum_{d \in [D]} \sum_{x \in \mathbb X^{k_d-1}} \left| \int_{\mathbb I^{r_<[k_d] \setminus \{1\}}} \left( \prod_{j=1}^{k_d-1} \mu^{N, \beta_j}(x_j) \right)
    \right.\right.\\&\hspace{3.5cm} \left.\left. 
    \cdot \left[ W^N_d(\alpha, \boldsymbol \beta) - W_d(\alpha, \boldsymbol \beta) \right] \, \mathrm d\boldsymbol \beta \right| \, \mathrm d\alpha \right] \\
    &\quad \leq |\mathbb X| M_f L_P|\mathbb X| \sum_{d \in [D]} \lVert W^N_d - W_d \rVert_{\square^{k_d-1}} \to 0
\end{align*}
by Assumption~\ref{ass:W}, and at rate $O(1/\sqrt N)$ if \eqref{eq:Wconv} converges at rate $O(1/\sqrt N)$.

For the third term, we have 
\begin{align*}
    &\E \left[ \left| \boldsymbol \mu^N_t P^{t, \boldsymbol \pi^N_t, \boldsymbol \mu^N_t}_{W}(f) - \boldsymbol \mu^N_t P^{t, \boldsymbol \pi_t, \boldsymbol \mu^N_t}_{W}(f) \right| \right] \\
    &\quad = \E \left[ \left| \int_{\mathbb I} \sum_{x \in \mathbb X} \mu^{N,\alpha}_t(x) \sum_{u \in \mathbb U} \left[ \pi^{N, \alpha}_t(u \mid x) - \pi^\alpha_t(u \mid x) \right] 
    \right.\right.\\&\qquad \left.\left. 
    \cdot \sum_{x' \in \mathbb X} P_t \left(x' \mid x, u, \nu^{\alpha, \boldsymbol \mu^N_t}_{W} \right) \, f(x', \alpha) \, \mathrm d\alpha \right| \right] \\
    &\quad \leq M_f \E \left[ \int_{\mathbb I} \left| \pi^{N, \alpha}_t(u \mid x) - \pi^\alpha_t(u \mid x) \right| \, \mathrm d\alpha \right] \\
    &\quad = M_f \E \left[ \sum_{j \in [N] \setminus \{i\}} \int_{I_j^N} \left| \pi^{\frac{\left\lceil N\alpha \right\rceil}{N}}_t(u \mid x) - \pi^\alpha_t(u \mid x) \right| \, \mathrm d\alpha \right] \\
    &\qquad + M_f \E \left[ \int_{I_i^N} \left| \hat \pi_t(u \mid x) - \pi^\alpha_t(u \mid x) \right| \, \mathrm d\alpha \right] \\
    &\quad \leq M_f \cdot \frac{L_\pi}{N} + M_f \cdot \frac{2|D_\pi|}{N} + M_f \cdot \frac{2}{N}
\end{align*}
by $\boldsymbol \pi \in \boldsymbol \Pi_\mathrm{Lip}$ with Lipschitz constant $L_\pi$ and up to $D_\pi$ discontinuities, where we bound the integrands by $2$. 

For the fourth term, we find that
\begin{align*}
    &\E \left[ \left| \boldsymbol \mu^N_t P^{t, \boldsymbol \pi_t, \boldsymbol \mu^N_t}_{W}(f) - \boldsymbol \mu^N_t P^{t, \boldsymbol \pi_t, \boldsymbol \mu_t}_{W}(f) \right| \right] \\
    &\quad = \E \left[ \left| \int_{\mathbb I} \sum_{x \in \mathbb X} \mu^{N,\alpha}_t(x) \sum_{u \in \mathbb U} \pi^\alpha_t(u \mid x) \sum_{x' \in \mathbb X} f(x', \alpha)
    \right.\right.\\&\quad \left.\left. 
    \cdot \left[ P_t \left(x' \mid x, u, \nu^{\alpha, \boldsymbol \mu^N_t}_{W} \right) - P_t \left(x' \mid x, u, \nu^{\alpha, \boldsymbol \mu_t}_{W} \right) \right] \, \mathrm d\alpha \right| \right] \\
    &\quad \leq |\mathbb X| M_f L_P\E \left[ \int_{\mathbb I} \left\Vert \nu^{\alpha, \boldsymbol \mu^N_t}_{W} - \nu^{\alpha, \boldsymbol \mu_t}_{W} \right\Vert \, \mathrm d\alpha \right] \\
    &\quad \leq |\mathbb X| M_f L_P\E \left[ \int_{\mathbb I} \sum_{d \in [D]} \sum_{x \in \mathbb X^{k_d-1}} \left| \int_{\mathbb I^{r_<[k_d] \setminus \{1\}}} W_d(\alpha, \boldsymbol \beta)
    \right.\right.\\&\hspace{2.5cm} \left.\left. 
    \cdot \left[ \prod_{j=1}^{k_d-1} \mu^{N, \beta_j}(x_j) - \prod_{j=1}^{k_d-1} \mu^{\beta_j}(x_j) \right] \, \mathrm d\boldsymbol \beta \right| \, \mathrm d\alpha \right] \\
    &\quad = |\mathbb X| M_f L_P\int_{\mathbb I} \sum_{d \in [D]} \sum_{x \in \mathbb X^{k_d-1}} \E \left[ \left| \int_{\mathbb I^{[k_d] \setminus \{1\}}} \int W_d(\alpha, \boldsymbol \beta, \boldsymbol \zeta) \, \mathrm d\boldsymbol \zeta
    \right.\right.\\&\hspace{2.5cm} \left.\left. 
    \cdot \left[ \prod_{j=1}^{k_d-1} \mu^{N, \beta_j}(x_j) - \prod_{j=1}^{k_d-1} \mu^{\beta_j}(x_j) \right] \, \mathrm d\boldsymbol \beta \right| \, \mathrm d\alpha \right] \\
    &\quad = |\mathbb X| M_f L_P\int_{\mathbb I} \sum_{d \in [D]} \sum_{x \in \mathbb X^{k_d-1}} 
    \\&\hspace{2cm}
    \E \left[ \left| \bigotimes^{k_d-1} \boldsymbol \mu^N_t(f'_{x, \alpha}) - \bigotimes^{k_d-1} \boldsymbol \mu_t(f'_{x, \alpha}) \right| \right] \, \mathrm d\alpha \to 0
\end{align*}
at the rate in the induction assumption, by applying the induction assumption \eqref{eq:muconv} for $n=k_d-1$ to the functions
\begin{align*}
    f'_{x, \alpha}(x', \beta) = \int_{\mathbb I^{r_<[k_d] \setminus [k_d]}} W_d(\alpha, \boldsymbol \beta, \boldsymbol \zeta) \, \mathrm d\boldsymbol \zeta \cdot \mathbbm 1_{\{x\}}(x')
\end{align*} 
for any $(x, \alpha) \in \mathbb X^{k_d-1} \times \mathbb I$.

For the fifth term, we analogously obtain
\begin{align*}
    &\E \left[ \left| \boldsymbol \mu^N_t P^{t, \boldsymbol \pi_t, \boldsymbol \mu_t}_{W}(f) - \boldsymbol \mu_t P^{t, \boldsymbol \pi_t, \boldsymbol \mu_t}_{W}(f) \right| \right] \\
    &\quad = \E \left[ \left| \int_{\mathbb I} \sum_{x \in \mathbb X} \left[ \mu^{N,\alpha}_t(x) - \mu^\alpha_t(x) \right] \sum_{u \in \mathbb U} \pi^\alpha_t(u \mid x) 
    \right.\right.\\&\hspace{1.5cm} \left.\left. 
    \cdot \sum_{x' \in \mathbb X} P_t \left(x' \mid x, u, \nu^{\alpha, \boldsymbol \mu_t}_{W} \right) \, f(x', \alpha) \, \mathrm d\alpha \right| \right] \\
    &\quad = \E \left[ \left| \boldsymbol \mu^N_t(f') - \boldsymbol \mu_t(f') \right| \right] \to 0 .
\end{align*}
at the rate in the induction assumption, by applying the induction assumption \eqref{eq:muconv} to
\begin{align*}
    f'(x, \alpha) = \sum_{u \in \mathbb U} \pi^\alpha_t(u \mid x) \sum_{x' \in \mathbb X} P_t \left(x' \mid x, u, \nu^{\alpha, \boldsymbol \mu_t}_{W} \right) \, f(x', \alpha) .
\end{align*}
This concludes the proof by induction.
\end{proof}

\subsection{Proof of Theorem~\ref{thm:xconv}}
The proof of Theorem~\ref{thm:xconv} mirrors the proof in \citet{cui2022learning} apart from propagating the multidimensional convergence results forward, and we give the entire proof for completeness and convenience. Again, we introduce some notation to improve readability. For any $\alpha \in \mathbb I$, $d \in [D]$, define maps $\nu^{\alpha}_{d} \colon \mathcal P(\mathbb X)^{\mathbb I} \to \mathcal P(\mathbb X)$ and $\nu^{\alpha}_{N,d} \colon \mathcal P(\mathbb X)^{\mathbb I} \to \mathcal P(\mathbb X)$ as
\begin{align*}
    \nu^{\alpha}_{d}(\boldsymbol \mu) &\coloneqq \int_{\mathbb I^{r_<[k_d] \setminus \{1\}}} W_d(\alpha, \boldsymbol \beta) \prod_{j=1}^{k_d-1} \mu^{\beta_j}(\cdot) \, \mathrm d\boldsymbol \beta, \\ 
    \nu^{\alpha}_{N,d}(\boldsymbol \mu) &\coloneqq \int_{\mathbb I^{r_<[k_d] \setminus \{1\}}} W^N_d(\alpha, \boldsymbol \beta) \prod_{j=1}^{k_d-1} \mu^{\beta_j}(\cdot) \, \mathrm d\boldsymbol \beta
\end{align*}
with $D$-dimensional shorthands
\begin{align*}
    \nu^{\alpha}(\boldsymbol \mu) &\coloneqq (\nu^{\alpha}_{d}(\boldsymbol \mu))_{d \in [D]}, \\ 
    \nu^{\alpha}_N(\boldsymbol \mu) &\coloneqq (\nu^{\alpha}_{N,d}(\boldsymbol \mu))_{d \in [D]}
\end{align*} 
such that by definition $\nu^\alpha_t = \nu^{\alpha}(\boldsymbol \mu_t)$ and $\nu^{N,i}_t = \nu^{\frac i N}_N(\boldsymbol \mu^N_t)$. 

\begin{proof}
To begin, we prove \eqref{eq:xconv} $\implies$ \eqref{eq:xmuconv} at any fixed time $t$. Define the uniform bound $M_h$ and uniform Lipschitz constant $L_h$ of functions in $\mathbb H$. For any $h \in \mathbb H$ we have
\begin{align*}
    &\left| \E \left[ h(X^i_{t}, \nu^{\frac i N}_N(\boldsymbol \mu^N_t)) \right] - \E \left[ h(\hat X^{\frac i N}_{t}, \nu^{\frac i N}(\boldsymbol \mu_t)) \right] \right| \\
    &\quad = \left| \E \left[ h(X^i_{t}, \nu^{\frac i N}_N(\boldsymbol \mu^N_t)) \right] - \E \left[ h(X^i_{t}, \nu^{\frac i N}_N(\boldsymbol \mu_t)) \right] \right| \\
    &\qquad + \left| \E \left[ h(X^i_{t}, \nu^{\frac i N}_N(\boldsymbol \mu_t)) \right] - \E \left[ h(X^i_{t}, \nu^{\frac i N}(\boldsymbol \mu_t)) \right] \right| \\
    &\qquad + \left| \E \left[ h(X^i_{t}, \nu^{\frac i N}(\boldsymbol \mu_t)) \right] - \E \left[ h(\hat X^{\frac i N}_{t}, \nu^{\frac i N}(\boldsymbol \mu_t)) \right] \right|
\end{align*}
which we will analyze as $N \to \infty$.

For the first term, we obtain
\begin{align*}
    &\left| \E \left[ h(X^i_{t}, \nu^{\frac i N}_N(\boldsymbol \mu^N_t)) \right] - \E \left[ h(X^i_{t}, \nu^{\frac i N}_N(\boldsymbol \mu_t)) \right] \right| \\
    &\quad \leq \E \left[ \E \left[ \left| h(X^i_{t}, \nu^{\frac i N}_N(\boldsymbol \mu^N_t)) - h(X^i_{t}, \nu^{\frac i N}_N(\boldsymbol \mu_t)) \right| \innermid X^i_{t} \right] \right] \\
    &\quad \leq L_h \E \left[ \left\Vert \nu^{\frac i N}_N(\boldsymbol \mu^N_t) - \nu^{\frac i N}_N(\boldsymbol \mu_t) \right\Vert \right] \\
    &\quad = L_h \E \left[ \sum_{d \in [D]} \sum_{x \in \mathbb X^{k_d-1}} \left| \int_{\mathbb I^{r_<[k_d] \setminus \{1\}}} W^N_d(\alpha, \boldsymbol \beta)
    \right.\right.\\&\hspace{2cm}\left.\left.
    \cdot \left[ \prod_{j=1}^{k_d-1} \mu^{N,\beta_j}(x_j) - \prod_{j=1}^{k_d-1} \mu^{\beta_j}(x_j) \right] \, \mathrm d\beta \right| \right] \to 0
\end{align*}
uniformly by applying Theorem~\ref{thm:muconv} to the functions
\begin{align*}
    f_{N,i,x}'(x', \beta) = \int_{\mathbb I^{r_<[k_d] \setminus [k_d]}} W^N_d(\frac i N, \boldsymbol \beta, \boldsymbol \zeta) \, \mathrm d\boldsymbol \zeta \cdot \mathbbm 1_{\{x\}}(x').
\end{align*}

For the second term, we analogously have
\begin{align*}
    &\left| \E \left[ h(X^i_{t}, \nu^{\frac i N}_N(\boldsymbol \mu_t)) \right] - \E \left[ h(X^i_{t}, \nu^{\frac i N}(\boldsymbol \mu_t)) \right] \right| \\
    &\quad \leq L_h \lVert \nu^{\frac i N}_N(\boldsymbol \mu_t) - \nu^{\frac i N}(\boldsymbol \mu_t) \rVert_1 \\
    &\quad \leq L_h \sum_{d \in [D]} \sum_{x \in \mathbb X^{k_d-1}} \left| \int_{\mathbb I^{r_<[k_d] \setminus \{1\}}} \left( \prod_{j=1}^{k_d-1} \mu^{\beta_j}(x_j) \right)
    \right.\\&\hspace{3.5cm} \left.
    \cdot \left[ W^N_d(\frac i N, \boldsymbol \beta) - W_d(\frac i N, \boldsymbol \beta) \right] \, \mathrm d\beta \right| \\
    &\quad \leq L_h \sum_{d \in [D]} \sum_{x \in \mathbb X^{k_d-1}} \left| \int_{\mathbb I^{r_<[k_d] \setminus \{1\}}} \left( \prod_{j=1}^{k_d-1} \mu^{\beta_j}(x_j) \right)
    \right.\\&\hspace{2.5cm} \left.
    \cdot \left[ W^N_d(\frac i N, \boldsymbol \beta) - N \int_{I_i^N} W_d(\alpha, \beta) \, \mathrm d\alpha \right] \, \mathrm d\beta \right| \\
    &\qquad + L_h \sum_{d \in [D]} \sum_{x \in \mathbb X^{k_d-1}} \left| \int_{\mathbb I^{r_<[k_d] \setminus \{1\}}} \left( \prod_{j=1}^{k_d-1} \mu^{\beta_j}(x_j) \right)
    \right.\\&\hspace{2.5cm} \left.
    \cdot \left[ N \int_{I_i^N} W_d(\alpha, \beta) \, \mathrm d\alpha - W_d(\frac i N, \boldsymbol \beta) \right] \, \mathrm d\beta \right|
\end{align*}
where for the former (finite) sum we have
\begin{align*}
    &\left| \int_{\mathbb I^{r_<[k_d] \setminus \{1\}}} \left( \prod_{j=1}^{k_d-1} \mu^{\beta_j}(x_j) \right)
    \right.\\&\hspace{2.5cm} \left.
    \cdot \left[ W^N_d(\frac i N, \boldsymbol \beta) - N \int_{I_i^N} W_d(\alpha, \beta) \, \mathrm d\alpha \right] \, \mathrm d\beta \right| \\
    &\quad = \left| N \int_{I_i^N} \int_{\mathbb I^{r_<[k_d] \setminus \{1\}}} \left( \prod_{j=1}^{k_d-1} \mu^{\beta_j}(x_j) \right)
    \right.\\&\hspace{2.5cm} \left.
    \cdot \left[ W^N_d(\alpha, \boldsymbol \beta) - W_d(\alpha, \beta) \right] \, \mathrm d\beta \, \mathrm d\alpha \right| \\
    &\quad \leq N \int_{I_i^N} \left| \int_{\mathbb I^{r_<[k_d] \setminus \{1\}}} \left( \prod_{j=1}^{k_d-1} \mu^{\beta_j}(x_j) \right)
    \right.\\&\hspace{2.5cm} \left.
    \cdot \left[ W^N_d(\alpha, \boldsymbol \beta) - W_d(\alpha, \beta) \right] \, \mathrm d\beta \right| \, \mathrm d\alpha \\
    &\quad \eqqcolon I_i^N
\end{align*}
since by definition of the step-hypergraphon, $W^N_d(\frac i N, \boldsymbol \beta) = W^N_d(\alpha, \boldsymbol \beta)$ over $\alpha \in I_i^N$. Therefore,
\begin{align*}
    \frac 1 N \sum_{i=1}^N I_i^N &= \int_{\mathbb I^{r_<[k_d]}} \left[ W^N_d(\beta) - W_d(\beta) \right] \prod_{j=1}^{k_d-1} \mu^{\beta_j}(x_j) \, \mathrm d\beta \to 0
\end{align*}
as in the proof of Theorem~\ref{thm:muconv} by Assumption~\ref{ass:W}. Fix $\varepsilon, p > 0$. As $N$ becomes sufficiently large, there must exist $\mathbb J^N_1$, $|\mathbb J^N_1| \geq \left\lfloor (1-p) N \right\rfloor$ such that
\begin{align*}
    I_i^N < \varepsilon, \quad \forall i \in \mathbb J^N_1 .
\end{align*}
We prove this by contradiction: Assume there does not exist such $\mathbb J^N_1$, then there exist at least $\left\lceil pN \right\rceil$ agents where $I_i^N \geq \varepsilon$. Since $I_i^N \geq 0$, it follows that $\frac 1 N \sum_{i=1}^N I_i^N \geq \frac 1 N \left\lceil pN \right\rceil \varepsilon \geq \varepsilon p$, which contradicts the convergence to zero of $\frac 1 N \sum_{i=1}^N I_i^N$. Repeating the argument for each $d \in [D]$, $x \in \mathbb X^{k_d-1}$ bounds the first sum.

For the latter (finite) sum, we have
\begin{align*}
    &\left| \int_{\mathbb I^{r_<[k_d] \setminus \{1\}}} \left( \prod_{j=1}^{k_d-1} \mu^{\beta_j}(x_j) \right)
    \right.\\&\hspace{2.5cm} \left.
    \cdot \left[ N \int_{I_i^N} W_d(\alpha, \beta) \, \mathrm d\alpha - W_d(\frac i N, \boldsymbol \beta) \right] \, \mathrm d\beta \right| \\
    &\quad = \left| N \int_{I_i^N} \int_{\mathbb I^{r_<[k_d] \setminus \{1\}}} \left( \prod_{j=1}^{k_d-1} \mu^{\beta_j}(x_j) \right)
    \right.\\&\hspace{2.5cm} \left.
    \cdot \left[ W_d(\alpha, \boldsymbol \beta) - W_d(\frac{\left\lceil N\alpha \right\rceil}{N}, \beta) \right] \, \mathrm d\beta \, \mathrm d\alpha \right| \\
    &\quad \leq N \int_{I_i^N} \int_{\mathbb I^{r_<[k_d] \setminus \{1\}}} \left| W_d(\alpha, \boldsymbol \beta) - W_d(\frac{\left\lceil N\alpha \right\rceil}{N}, \beta) \right| \, \mathrm d\beta \, \mathrm d\alpha \\
    &\quad \leq N \frac 1 N \cdot \frac{L_W}{N} = \frac{L_W}{N} \to 0
\end{align*}
by Assumption~\ref{ass:Lip}. Alternatively, under only block-wise Lipschitz $W$ as in \eqref{eq:blockwiseLip}, the same result is obtained by first separating out finitely many $i$ (at most $Q-1$) for which Lipschitzness fails, trivially bounding their terms by $\frac{2(Q-1)}{N}$. For all other $i$, there exists $k \in \{1, \ldots, Q\}$ such that $I_i^N \times I_j \subseteq I_k \times I_j$, i.e. the Lipschitz bound applies.

For the third term, again fix $\varepsilon, p > 0$. Then, by our initial assumption of \eqref{eq:xconv}, for sufficiently large $N$ there exists a set $\mathbb J^N_2$, $|\mathbb J^N_2| \geq \left\lfloor (1-p) N \right\rfloor$ such that
\begin{align*}
    &\left| \E \left[ h(X^i_{t}, \nu^{\frac i N}(\boldsymbol \mu_t)) \right] - \E \left[ h(\hat X^{\frac i N}_{t}, \nu^{\frac i N}(\boldsymbol \mu_t)) \right] \right| < \varepsilon, \quad \forall i \in \mathbb J^N_2
\end{align*}
independent of $\hat \pi \in \Pi$.

This completes the proof of \eqref{eq:xconv} $\implies$ \eqref{eq:xmuconv} at any time $t$, since by the prequel, the intersection of all correspondingly chosen, finitely many sets $\mathbb J^N_i$ for sufficiently large $N$ has at least $N - \sum_i \left\lceil p_i N \right\rceil$ elements, which is always larger than $N - \left\lceil p N \right\rceil$ for any $p>0$ by choosing $p_i$ sufficiently small.

Finally, we show \eqref{eq:xconv} at all times $t$ using the prequel by induction, which will imply \eqref{eq:xmuconv}. By definition for $t=0$, $\hat X^{\frac i N}_{t} \sim \mu_0$ and $X^i_{t} \sim \mu_0$ imply
\begin{align*}
    \left| \E \left[ g(X^i_{0}) \right] - \E \left[ g(\hat X^{\frac i N}_{0}) \right] \right| = 0.
\end{align*}

For the induction step, define the uniform bound $M_g$ of functions in $\mathbb G$. Observe that
\begin{align*}
    &\left| \E \left[ g(X^i_{t+1}) \right] - \E \left[ g(\hat X^{\frac i N}_{t+1}) \right] \right| \\
    &\quad = \left| \E \left[ l_{N,t}(X^i_{t}, \nu^{\frac i N}_N(\boldsymbol \mu^N_t)) \right] - \E \left[ l_{N,t}(\hat X^{\frac i N}_{t}, \nu^{\frac i N}(\boldsymbol \mu_t)) \right] \right|
\end{align*}
using the uniformly bounded and Lipschitz functions
\begin{align*}
    l_{N,t}(x, \nu) \coloneqq \sum_{u \in \mathbb U} \hat \pi_t(u \mid x) \sum_{x' \in \mathbb X} P_t(x' \mid x, u, \nu) g(x')
\end{align*}
with bound $M_g$ and Lipschitz constant $|\mathbb X| M_g L_P$. By induction assumption \eqref{eq:xconv} and \eqref{eq:xconv} $\implies$ \eqref{eq:xmuconv}, there exists $N' \in \mathbb N$ such that for all $N > N'$ we have
\begin{align*}
    \left| \E \left[ l_{N,t}(X^i_{t}, \nu^{\frac i N}_N(\boldsymbol \mu^N_t)) \right] - \E \left[ l_{N,t}(\hat X^{\frac i N}_{t}, \nu^{\frac i N}(\boldsymbol \mu_t)) \right] \right| < \varepsilon
\end{align*}
uniformly over $\hat \pi \in \Pi, i \in \mathbb J^N$ for some $\mathbb J^N \subseteq [N]$ with $|\mathbb J^N| \geq \left\lfloor (1-p) N \right\rfloor$. This concludes the proof by induction.
\end{proof}

\subsection{Proof of Corollary~\ref{coro:jconv}}
\begin{proof}
The result follows more or less directly from Theorem~\ref{thm:xconv}. Consider first the finite horizon case $\mathbb T = \{0, 1, \ldots, T-1\}$. Define
\begin{align*}
    R_t^{\hat \pi}(x, \nu) \coloneqq \sum_{u \in \mathbb U} R_t(x,u,\nu) \hat \pi_t(u \mid s)
\end{align*}
with uniform bound $M_R$ and Lipschitz constant $|U| L_R$. Therefore, by choosing the maximum over all $N'$ for all finitely many times $t \in \mathbb T$ via Theorem~\ref{thm:xconv}, there exists $N' \in \mathbb N$ such that for all $N > N'$ we have
\begin{align*}
    &\left| J_i^N(\pi^1, \ldots, \pi^{i-1}, \hat \pi, \pi^{i+1}, \ldots, \hat \pi) - J^{\boldsymbol \mu}_{\frac i N}(\hat \pi) \right| \\
    &\quad \leq \sum_{t=0}^{T-1} \left| \E \left[ R_t^{\hat \pi_t}(X_t^i,  \nu^{\frac i N}(\boldsymbol \mu_t)) \right] - \E \left[ R_t^{\hat \pi_t}(\hat X_t^{\frac i N}, \nu^{\frac i N}(\boldsymbol \mu_t)) \right] \right| < \varepsilon .
\end{align*}
uniformly over $\hat \pi \in \Pi, i \in \mathbb J^N$ for some $\mathbb J^N \subseteq [N]$ with $|\mathbb J^N| \geq \left\lfloor (1-p) N \right\rfloor$.

For the infinite horizon problem $\mathbb T = \mathbb N_0$, we first pick some time $T' > \frac {\log \frac {\varepsilon (1-\gamma)}{4M_R}} {\log \gamma}$ such that
\begin{align*}
    &\sum_{t=0}^{T'-1} \gamma^t \left| \E \left[ r_{\hat \pi_t}(X_t^i,  \nu^{\frac i N}(\boldsymbol \mu_t)) \right] - \E \left[ r_{\hat \pi_t}(\hat X_t^{\frac i N}, \nu^{\frac i N}(\boldsymbol \mu_t)) \right] \right| \\
    &+ \gamma^{T'} \sum_{t=T'}^\infty \gamma^{t-T'} \left| \E \left[ r_{\hat \pi_t}(X_t^i,  \nu^{\frac i N}(\boldsymbol \mu_t)) \right] - \E \left[ r_{\hat \pi_t}(\hat X_t^{\frac i N}, \nu^{\frac i N}(\boldsymbol \mu_t)) \right] \right| \\
    &< \sum_{t=0}^{T'-1} \gamma^t \left| \E \left[ r_{\hat \pi_t}(X_t^i,  \nu^{\frac i N}(\boldsymbol \mu_t)) \right] - \E \left[ r_{\hat \pi_t}(\hat X_t^{\frac i N}, \nu^{\frac i N}(\boldsymbol \mu_t)) \right] \right| + \frac \varepsilon 2
\end{align*}
and again apply Theorem~\ref{thm:xconv} to the remaining finite sum.
\end{proof}

\subsection{Proof of Corollary~\ref{coro:approxnash}}
\begin{proof}
The result follows directly from Corollary~\ref{coro:jconv}. Let $\varepsilon > 0$, then by Corollary~\ref{coro:jconv} there exists $N' \in \mathbb N$ such that for all $N > N'$ we have
\begin{align*}
    &\max_{\pi \in \Pi} \left( J_i^N({\pi^1}, \ldots, {\pi^{i-1}}, \pi, {\pi^{i+1}}, \ldots, {\pi^N}) - J_i^N(\pi^1, \ldots, \pi^N) \right) \\
    &\quad \leq \max_{\pi \in \Pi} \left( J_i^N({\pi^1}, \ldots, {\pi^{i-1}}, \pi, {\pi^{i+1}}, \ldots, {\pi^N}) - J^{\boldsymbol \mu}_{\frac i N}(\pi) \right) \\
    &\qquad + \max_{\pi \in \Pi} \left( J^{\boldsymbol \mu}_{\frac i N}(\pi) - J^{\boldsymbol \mu}_{\frac i N}(\pi^{\frac i N}) \right) \\
    &\qquad + \left( J^{\boldsymbol \mu}_{\frac i N}(\pi^{\frac i N}) - J_i^N(\pi^1, \ldots, \pi^N) \right) \\
    &\quad < \frac{\varepsilon}{2} + 0 + \frac{\varepsilon}{2} = \varepsilon
\end{align*}
uniformly over $i \in \mathbb J^N$ for some $\mathbb J^N \subseteq [N]$ with $|\mathbb J^N| \geq \left\lfloor (1-p) N \right\rfloor$, where by definition of equilibrium optimality we obtained
\begin{align*}
    \max_{\pi \in \Pi} \left( J^{\boldsymbol \mu}_{\frac i N}(\pi) - J^{\boldsymbol \mu}_{\frac i N}(\pi^{\frac i N}) \right) = 0.
\end{align*}
This concludes the proof.
\end{proof}

\begin{figure*}
    \centering
    \includegraphics[width=\linewidth]{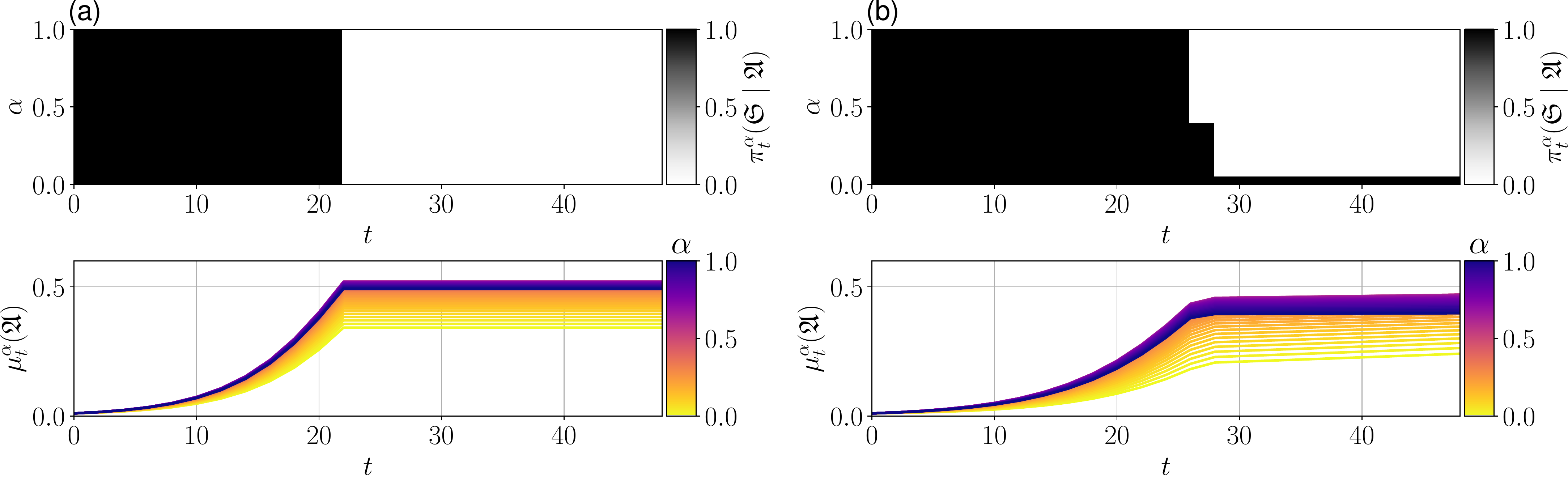}
    \caption{Analysis of equilibrium behavior for the Rumor problem on additional hypergraphon configurations. (a): $(W_\mathrm{rank}, \hat W_\mathrm{inv-unif})$; (b): $(W_\mathrm{unif}, \hat W_\mathrm{inv-unif})$.}
    \label{fig:rumormore}
\end{figure*}

\begin{figure*}
    \centering
    \includegraphics[width=\linewidth]{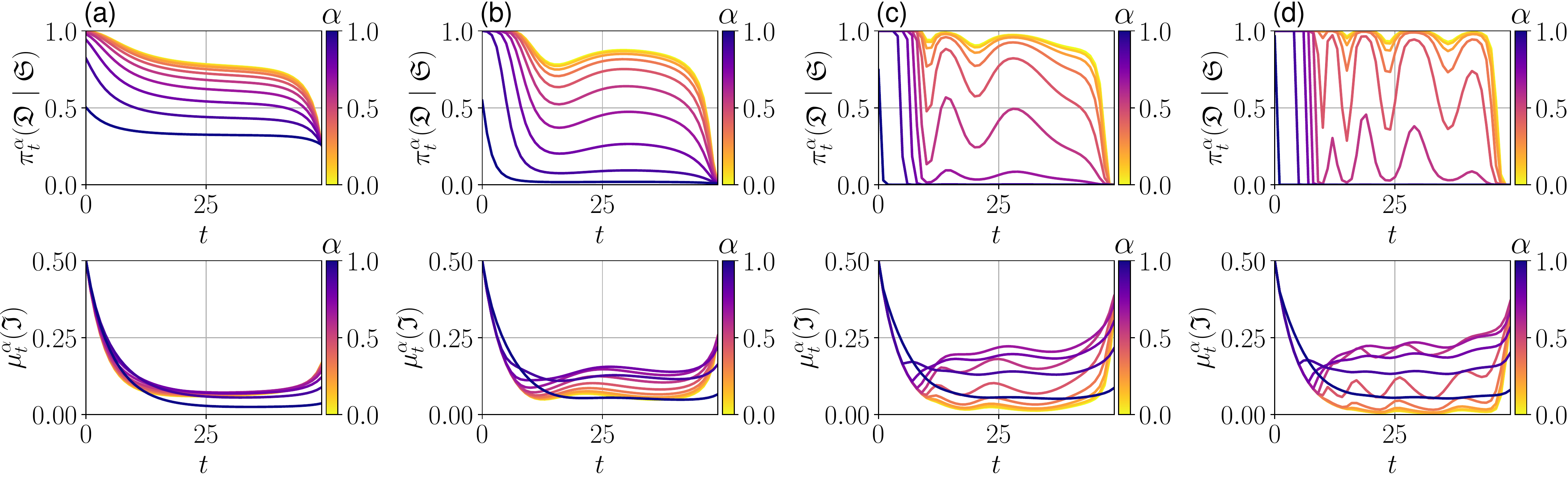}
    \caption{The solution policy and corresponding mean field for graphons $(W_\mathrm{unif}, \hat W_\mathrm{unif})$ from Fig.~\ref{fig:sis_equilibrium} at different iterations $n$. It can be observed that in the SIS problem, the solution oscillates between taking precautions and not taking precautions. (a): $n=20$; (b): $n=100$; (c): $n=500$; (d): $n=1500$.}
    \label{fig:sis_oscillation}
\end{figure*}

\begin{figure}
    \centering
    \includegraphics[width=\linewidth]{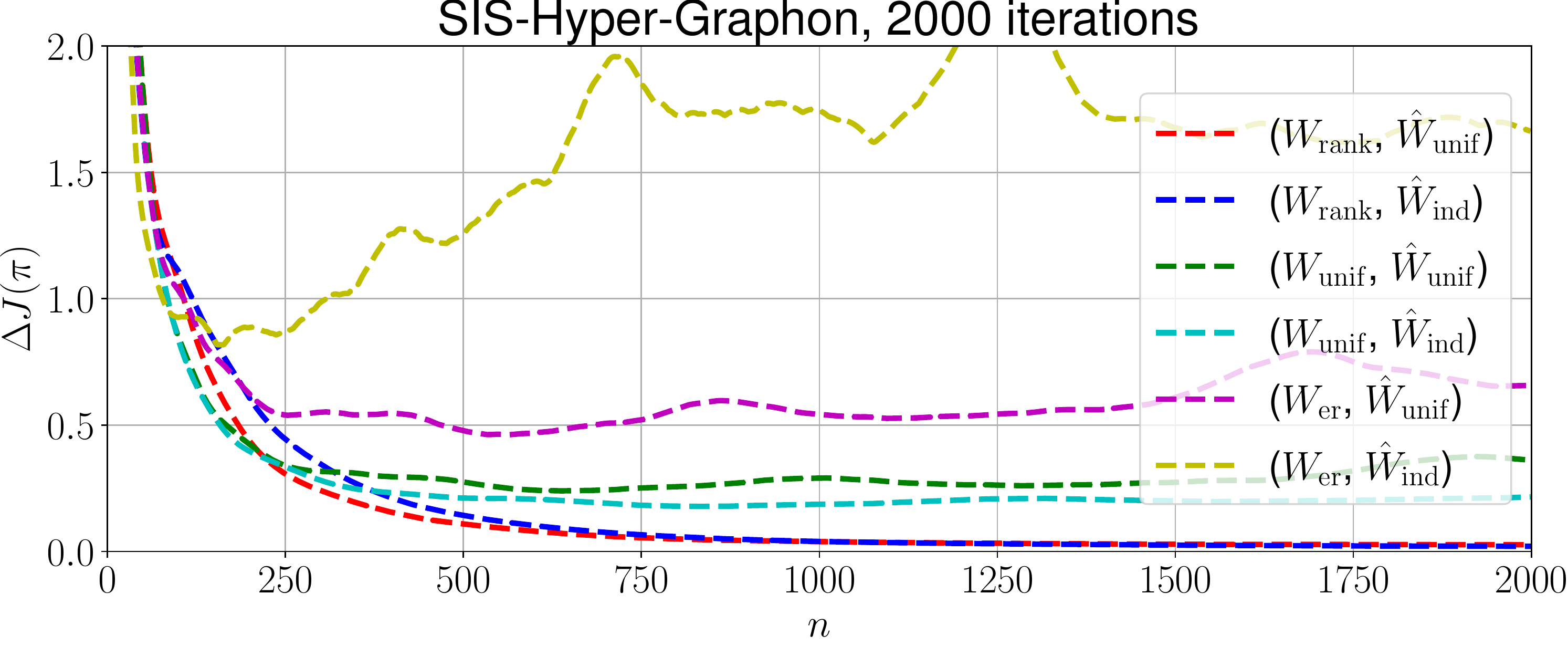}
    \caption{Average exploitability over iterations $n$ of Online Mirror Descent \cite{perolat2021scaling} on the SIS problem. It can be observed that for some configurations, the method fails to converge to an equilibrium.}
    \label{fig:sis_equilibrium}
\end{figure}

\section{Additional experiments \label{app:exp}}
In Fig.~\ref{fig:rumormore}, we show additional results for the Rumor problem and inverted $3$-uniform hypergraphons. There, we find almost inverted results as in Fig.~\ref{fig:rumor}, indicating that the influence of connections from the second layer are more important under the given problem parameters. However, we note that surprisingly, the highest awareness is reached for intermediate $\alpha$.

As an additional example, in the timely SIS problem, we assume that there exists an epidemic that spreads to neighboring nodes according to the classical SIS dynamics, see e.g. \citet{kiss2017mathematics}. Analogously, we may consider extensions to arbitrary variations of the SIS model such as SIR or SEIR. Each healthy (or susceptible, $\mathfrak S$) agent can take costly precautions ($\mathfrak P$) to avoid becoming infected ($\mathfrak I$), or ignore ($\bar{\mathfrak P}$) precautions at no further cost. Since being infected itself is costly, an equilibrium solution must balance the expected cost of infections against the cost of taking precautions. 

Formally, we define the state space $\mathbb X = \{\mathfrak S, \mathfrak I\}$ and action space $\mathbb U = \{\bar{\mathfrak P}, \mathfrak P\}$ such that
\begin{align*}
    P(\mathfrak I \mid \mathfrak S, \bar{\mathfrak P}, \boldsymbol \nu) &= \min \left( 1, \sum_{d \in [D]} \tau_d \boldsymbol \nu_d \left( \mathbbm 1_{\{\mathfrak I\}} \right) \right)\\
    P(\mathfrak I \mid \mathfrak S, \mathfrak P, \cdot) &= 0, \quad P(\mathfrak S \mid \mathfrak I, \cdot, \cdot) = \delta
\end{align*}
with infection rates $\tau_d > 0$, $\sum_d \tau_d \leq 1$, recovery rate $\delta \in (0,1)$ and rewards $R(x, u, \cdot) = c_P \mathbbm 1_{\{\mathfrak P\}}(u) + c_I \mathbbm 1_{\{\mathfrak I\}}(x)$ with infection and precaution costs $c_P > 0$, $c_I > 0$. In our experiments, we will use $\tau_d = 0.8$, $\delta = 0.2$, $c_P = 0.5$, $c_I = 2$, $\mu_0(\mathfrak I) = 0.5$ and $\mathbb T = \{0, 1, \ldots, 49\}$.

Existing state-of-the-art approaches such as online mirror descent (OMD) \cite{perolat2021scaling}(and similarly fictitious play, see e.g. \citet{cui2021approximately}) as depicted in Fig.~\ref{fig:sis_oscillation} and Fig.~\ref{fig:sis_equilibrium} for $10$ discretization points did not converge to an equilibrium in the considered $2000$ iterations, though we expect that the methods will converge when running for significantly more iterations -- e.g. $400000$ iterations as in \cite{lauriere2022scalable} -- which we could not verify here due to the computational complexity. We expect that existing standard results using monotonicity conditions \cite{lasry2007mean, perolat2021scaling} can be extended to the hypergraphon case in order to guarantee convergence of aforementioned learning algorithms. However, this remains outside the scope of our work. In particular for the ranked-attachment graphon and hypergraphon, the final behavior as seen in Fig.~\ref{fig:sis_oscillation} remains with an average final exploitability $\Delta J$ of above $0.25$, which is defined as
\begin{align*}
    \Delta J(\pi) = \int_{\mathbb I} \sup_{\pi^* \in \Pi} J^{\Psi(\pi)}_\alpha(\pi^*) - J^{\Psi(\pi)}_\alpha(\pi) \, \mathrm d\alpha
\end{align*}
and must be zero for an exact equilibrium.

\section*{References}
\bibliography{aipsamp}

\end{document}